\documentclass[reprint,
amsmath,amssymb,aps,
pre,showkeys,
floatfix]{revtex4-2}
\usepackage{dcolumn,bm,chemformula,graphicx,titlesec}
\usepackage[unicode]{hyperref}
\hypersetup{
     colorlinks= true,
     citecolor= blue,
     linkcolor= red,
     urlcolor=magenta,
     filecolor=cyan,
     linkbordercolor={1 0 0},
     citebordercolor={0 1 0},
     urlbordercolor={0 1 1},
     breaklinks=true, 			
} 
\usepackage[mathlines]{lineno}
\usepackage[section]{placeins}
\usepackage{float,wrapfig}
\usepackage[tight,TABTOPCAP]{subfigure}
\DeclareUnicodeCharacter{2212}{-}
\begin{document}
\preprint{APS/123-QED}
\title{Energetic and entropic cost due to overlapping of Turing-Hopf instabilities in presence of Cross Diffusion}
\author{Premashis Kumar}
\affiliation{S. N. Bose National Centre For Basic Sciences, Block-JD, Sector-III, Salt Lake, Kolkata 700 098, India}
\author{Gautam Gangopadhyay}
 \email{gautam@bose.res.in}
\affiliation{S. N. Bose National Centre For Basic Sciences, Block-JD, Sector-III, Salt Lake, Kolkata 700 098, India}
\date{\today}
\begin{abstract}
A systematic introduction to nonequilibrium thermodynamics of  dynamical instabilities are considered for an open nonlinear system beyond conventional Turing pattern in presence of cross diffusion. An altered condition of Turing instability in presence of cross diffusion can be best viewed in terms of critical control parameter and  wave number containing both the self and cross diffusion coefficients. Our main focus is on entropic and energetic cost of Turing-Hopf interplay in stationary pattern formation. Depending on the  relative dispositions of  Turing-Hopf  codimensional instabilities   from the reaction diffusion equation  it clarifies two aspects: energy cost of pattern formation, specially how Hopf instability can be utilized to dictate a stationary concentration profile, and the possibility of  revealing  nonequilibrium phase transition. In the Brusselator model  to understand these phenomena, we have analyzed through the relevant complex Ginzberg-Landau equation using multiscale  Krylov-Bogoiubov averaging method. Due to Hopf instability it is observed that the cross diffusion parameters can be a source of huge change in free energy and concentration profiles.
\end{abstract}

\maketitle

\section{\label{sec:level1}Introduction}

The traditional Turing pattern\citep{Cross2009PatternSystems, Turing1952TheMORPHOGENESIS, epstein1998introduction, castets1990experimental, ouyang1991transition} 
with very different self-diffusion coefficients and travelling waves\citep{zaikin1970concentration, winfree1972spiral, zhabotinsky1995pattern} are prevalent 
in the living tissues as morphogens\citep{Murray2003MathematicalApplications, kondo2010reaction, iber2013control, kretschmer2016pattern}, in cellular rhythms\citep{goldbeter1997biochemical,falcke2004reading,
thurley2012fundamental} and in many such situations\citep{Kondepudi2014ModernThermodynamics, Kuramoto1984ChemicalTurbulence} of a reaction-diffusion system and  can be drastically modified due to the slight presence of cross diffusion which is still under-investigated. For thermodynamics of pattern formation or more generally far from equilibrium system is addressed at length in the literature starting from the description of dissipative energy loss\citep{Prigogine1968SymmetryII, Nicolis1977Self-organizationFluctuations} to 
stochastic thermodynamics\citep{qian2006open, qian2016entropy}, along with thermal transport problems\citep{PhysRevEVanBroeck, Proesmans_2015} and  demonstrations of the validity of fluctuation theorems\citep{collin2005verification,Hummer3658}, paves the way to a systematic calculation of thermodynamic quantities in open dynamical systems. From a theoretical point of view it is still challenging to develop an approach to deal with an arbitrary nonlinear nonequilibrium process to tackle the problems of complex chemical 
network\citep{Polettini2014IrreversibleLaws,Rao2016NonequilibriumThermodynamics}  in a heterogeneous medium. Our goal here is to develop theories of nonquilibrium consequences of various dynamical instabilities in open systems describable as a reaction-diffusion system. Particularly the dynamical characterization of inbuilt limit cycle oscillation in presence of cross diffusion coefficients resulting  from diffusive flux of one species due to gradient in concentration of another\citep{Kondepudi2014ModernThermodynamics}, take its toll by altering their bifurcation scenario.

Whenever a closed system is opened by chemostatting, either a subset of conservation laws are broken or an emergent cycle appears for each chemostatted species\citep{Polettini2014IrreversibleLaws}. If there is no emergent cycle for open chemical reaction network with homogeneous chemostatting, then the system is said to be  unconditionally detailed-balanced for finite number of species and reactions due to absence of any non-conservative forces \citep{Rao2016NonequilibriumThermodynamics}. 
In open reaction diffusion system, Gibbs free energy is not minimized  due to the breaking of conservation laws which is characterized for closed system. Analogous to the definition of grand potential in terms of the Gibbs free energy  in equilibrium thermodynamics, the semigrand Gibbs free energy of the open system can be defined from the nonequilibrium Gibbs free energy of the closed system by subtracting the energetic contribution due to exchange of matter between chemostats and system \citep{Alberty2003ThermodynamicsReactions}. 
In reaction diffusion system, amplitude equation\citep{aranson2002world} is already used to capture a large degree of richness of pattern-formation both qualitatively and quantitatively near the onset of the instability\citep{Cross2009PatternSystems}. To treat the generic nonlinear dynamics with  symmetries and bifurcation characteristics of the system one can find the description of multiscale perturbation theory to obtain the amplitude equation\citep{Kuramoto1984ChemicalTurbulence, Walgraef1997ThePatterns, aranson2002world}, in terms of the complex Ginzburg Landau equation(CGLE).

In this context quantifying entropic and energetic costs of various pattern formation and interplay of various nonlinearity induced instabilities are of crucial theoretical concern here. This kind of approach is  adopted  recently in studying of the thermodynamics of Turing pattern in the presence of self diffusion only\cite{Falasco2018InformationPatterns} and chemical waves\cite{Avanzini2019ThermodynamicsWaves}. Again mathematical analysis of Turing-Hopf interplay has got some attention in different dynamical contexts\citep{DeWitA1996SpatiotemporalPoint., Just2001Spatiotemporal, Yang2003OscillatoryLayers, Ricard2009TuringBifurcation}, but the thermodynamic description of the overlap of Turing-Hopf instabilities is still missing. Moreover, in the study of pattern formation, very often cross diffusion coefficients of the species have been ignored, however, they can have very significant effect 
to modify almost all the patterns even if they are minimal\citep{vanagcross}. In  reaction diffusion system corresponding to traditional Turing pattern, threshold of Turing and Hopf instabilities are well-separated for very different diffusion coefficients of activator and inhibitor. Proper choice of cross diffusion coefficients can bring threshold of Turing and Hopf instabilities  close enough so that they eventually overlap and as a consequence a large variety of complex spatio-temporal pattern likely to emerge beyond critical Turing-Hopf point. As the usual multiscale methods\citep{Cross2009PatternSystems, Kuramoto1984ChemicalTurbulence} of deriving amplitude equation specially for reaction-diffusion system with cross-diffusion is rather cumbersome, we have employed here a simple method of derivation based on Krylov-Bogolyubov(KB) averaging method\citep{krylov1949introduction} to obtain the relevant Ginzberg-Landau equation.

The layout of the paper is as follows. In sec. \ref{sec:level2} we have discussed on chemostated Brusselator model with cross diffusion. Turing and Hopf instabilities are estimated for this system in the next section. In sec. \ref{sec:amplitudeequation} we have derived the amplitude equation  using Krylov-Bogolyubov method. Entropy production rate is calculated for the reaction-diffusion system in sec. \ref{sec:entropy}. In the next section nonequilibrium Gibbs free energy of chemostated system is formulated. In sec. \ref{confld} concentration fields of the intermediate species are obtained using analytical approach.  We have   provided numerical results and discussions in sec. \ref{RD}. Finally, the paper is concluded in Sec. \ref{con}.

\section{\label{sec:level2}Brusselator model with cross diffusion}

\begin{figure}[t]
\centering
    \includegraphics[width=\linewidth]{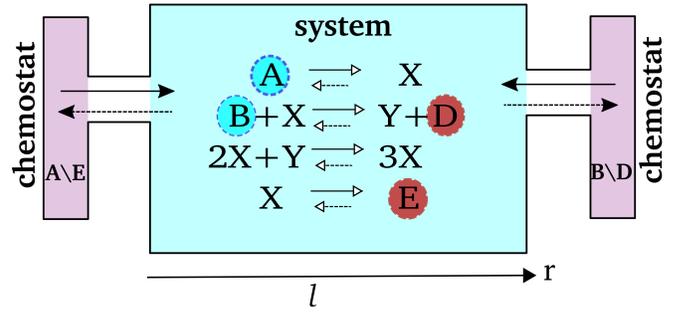}
\caption{\label{bm}Illustration of the Brusselator model as an open chemical network where A and B are reference chemostatted species. One can also use D and E as reference chemostatted species. Reservoirs of chemostatted species are shown in the two sides(purple color) and finite system of length $l$ is shown in the middle(sky blue).}
\end{figure}

The Brusselator model\citep{Prigogine1968SymmetryII, Nicolis1977Self-organizationFluctuations} is a prototype for studying various cooperative behavior in chemical kinetics and can successfully mimic oscillatory Belousov-Zhabotinsky reaction\citep{Zhabotinsky1991AWaves}. The reversible Brusselator model contains the following sequence of chemical reactions: 

\begin{equation}
\begin{aligned}
\rho&=1:&\ch{A&<=>[\text{k_1}][\text{k_{-1}}] X} \\
\rho&=2: &\ch{B + X&<=>[\text{k_{2}}][\text{k_{-2}}]Y + D}\\
\rho&=3:& \ch{2 X + Y&<=>[\text{k_{3}}][\text{k_{-3}}]3X} &\textsf{(Autocatalytic)} \\ 
\rho&=4:&\ch{X&<=>[\text{k_{4}}][\text{k_{-4}}]E}
 \end{aligned}
 \label{crn}
\end{equation}
where $'\rho'$ is reaction step label, $\{X,Y\}\in I$ are two intermediate species having dynamic concentration and  $\{A, B, D, E\}\in C$ are initial and final products with a constant homogeneous concentration along the entire system within the time scale of interest. Main features of the Brusselator model as an open chemical reaction network are presented in  FIG. \ref{bm}.

The stoichiometric matrix of the Brusselator  reaction network in eq. \eqref{crn} is
\begin{gather}
S_{\rho}^{\sigma}=
\bordermatrix{ ~ & R_{1} & R_{2}&R_{3}&R_{4}\cr
                 X&1 &-1&1&-1\cr
                 Y&0&1&-1&0 \cr
                 A&-1&0&0&0 \cr
                 B&0&-1&0&0\cr
                 D&0&1&0&0 \cr
                 E&0&0&0&1\cr} .\label{st}
 \end{gather}

With the assumptions, all the reverse  rate constants $k_{-\rho}$ are vanishingly small($10^{-4}$), and  the  forward  reaction  rate  constants $k_{\rho}$ are much higher than the reverse one, i.e., $k_{\rho}\gg k_{-\rho}$, the rate equations of concentrations of intermediate species in eq. \eqref{crn} read as
\begin{equation}
\begin{aligned}
  \dot{x}&={k_1}a-({k_2}b+k_4)x+{k_3}x^2y \\
  \dot{y}&={k_2}bx-{k_3}x^2y 
  \label{dynamic}
\end{aligned}
\end{equation}
where concentration of species are denoted by lowercase letters  $$x=[X],y=[Y],b=[B],a=[A] .$$
Steady-state value of the eq. \eqref{dynamic} that satisfies $\dot{x}=\dot{y}=0$ is,
$x_{0}=\frac{k_1}{k_4}a, y_{0}=\frac{{k_2}{k_4}}{{k_1}{k_3}}\frac{b}{a}$.

Now after taking diffusion into account, the reaction diffusion equation of the Brusselator model in one spatial dimension $r\in [0,l]$ could be specified as
\begin{equation}
\begin{aligned}
  \dot{x}&={k_1}a-({k_2}b+k_4)x+{k_3}x^2y+D_{11}x_{rr}+D_{12}y_{rr} \\
  \dot{y}&={k_2}bx-{k_3}x^2y+D_{21}x_{rr}+D_{22}y_{rr} 
  \label{ddynamic}
\end{aligned}
\end{equation}
in which $D_{11}$ , $D_{22}$ are self diffusion coefficients of intermediate species $X$ and $Y$ respectively and $D_{12}$ , $D_{21}$ are cross diffusion coefficients of $X$  and $Y$, respectively. 

This cross-diffusion coefficients generally have concentration dependence\cite{vanagcross}. Most importantly, the vanishingly low concentration of the species, ${\sigma}$ demands no flux of the species ${\sigma}(\sigma=1,2,..)$. Therefore cross-diffusion coefficients $D_{{\sigma}{\sigma}^{\prime}}({\sigma}\neq {\sigma}^{\prime})$ must tend to vanish as the concentration $z_{\sigma}$ tends to zero irrespective of the gradient in the concentration, $z_{{\sigma}^{\prime}}$. Following the work of Chung and Peacock-Lopez \cite{jessicalopez}, we can represent the concentration dependence of the cross diffusion coefficients as the 
\begin{equation}
D_{{\sigma}{\sigma}^{\prime}}(z_{\sigma})=\frac{{D_{{\sigma}{\sigma}^{\prime}}}z_{\sigma}}{\eta+z_{\sigma}}. \label{eq1}
\end{equation}
According to eq. \ref{eq1}, for $z_{\sigma}=0$, $D_{{\sigma}{\sigma}^{\prime}}$ will always vanish and thus, it satisfies the demand mentioned above. Moreover, for the minimal value of the $\eta$ with respect to concentrations  i.e., $\eta<<z_{\sigma}$, $D_{{\sigma}{\sigma}^{\prime}}(z_{\sigma})$ will be merely equal to the constant $D_{{\sigma}{\sigma}^{\prime}}$. Whereas, if the constant $\eta$  is very large compare to concentations i.e.  $\eta>>z_{\sigma}$, then $D_{{\sigma}{\sigma}^{\prime}}(z_{\sigma})=\frac{D_{{\sigma}{\sigma}^{\prime}}z_{\sigma}}{\eta}={D^\prime_{{\sigma}{\sigma}^{\prime}}} z_{\sigma}$ with ${D^\prime_{{\sigma}{\sigma}^{\prime}}}=\frac{D_{{\sigma}{\sigma}^{\prime}}}{\eta}$ and thus cross diffusion coefficients have linear concentration dependence. For simplicity, we have considered  here the case($\eta<<z_i$) of constant cross diffusion coefficients  for the most of the analysis for dynamic and thermodynamic entities in the Brusselator model.

\section{\label{thins}Turing and Hopf instabilities in the Brusselator model}

In reaction-diffusion system, one can have both Hopf and Turing instabilities which can be obtained from linear stability analysis. Evolution equations of the reaction diffusion system can be found by considering single Fourier mode of the form $\exp({\lambda(q)t+iqr}$) where growth rate has wave number, q dependence. For linear stability analysis at the steady-state value $(x_0,y_0)$, one needs the Jacobian matrix of the Brusselator model,
\begin{gather}
\mathcal{J}=
\begin{pmatrix}   
                 -({k_2}b+k_4)+2{k_3}x_0y_0 & {k_3}{x_0}^2 \cr
                 {k_2}b-2{k_3}x_0y_0& -{k_3}{x_0}^2\cr
\end{pmatrix}.\label{jaco}                 
\end{gather}
Here elements of the Jacobian matrix, $\mathcal{J}$ are the following

$J_{11}=-({k_2}b+k_4)+2{k_3}x_0y_0$, $J_{12}={k_3}{x_0}^2$

$J_{21}={k_2}b-2{k_3}x_0y_0$, $J_{22}=-{k_3}{x_0}^2.$

The Oregonator model\cite{fieldkoros, FieldNoyes}, the simplest model for describing the oscillations in BZ reaction have Jacobian with a sign structure(pure activator-inhibitor) opposite to that of the Brusselator model(cross activator-inhibitor).
\subsection{\label{turingins} Turing instability}
When the cross diffusion coefficients are present in the system and contribute to the Turing pattern, then self diffusion coefficients no need to obey the condition of local activation and lateral inhibition\citep{Murray2003MathematicalApplications}. Kumar and Horsthemke showed that presence of cross diffusion strongly modifies the Turing instability conditions and in this case Turing instability can arise even if the self diffusion coefficient of the inhibitor is more than the self diffusion of the activator\cite{Nkumar}. Zemskov et al. have presented universal conditions of the Turing instability in the presence of the cross diffusion coefficients with a linear concentration dependence. With the aid of those conditions, they have described the proper Turing instability region\cite{Zemskov}. Lin et al. have investigated the influence of cross diffusion in selecting the spatial pattern for the Busselator model in a three dimensional domain\cite{LIN2014}. By using finite volume element approximation, they have shown that cross diffusion can generate Turing pattern in this three-dimensional case. Exploiting conditions of Turing instability in the presence of cross diffusion, one can obtain a critical value of the control parameter and wave number . We would next find out those critical values in the 1D Brusselator model in the presence of both self and cross diffusion coefficients.      

In the presence of diffusion, Jacobian $\mathcal{J}$ becomes
\begin{gather}
\mathcal{J_D}=\mathcal{J}-q^2\mathcal{D}\nonumber\\
=\begin{pmatrix}   
                 -({k_2}b+k_4)+2{k_3}x_0y_0 & {k_3}{x_0}^2 \cr
                 {k_2}b-2{k_3}x_0y_0  & -{k_3}{x_0}^2 \cr
\end{pmatrix}
- q^2
\begin{pmatrix}
D_{11}&D_{12}\cr
D_{21}&D_{22}\cr
\end{pmatrix}
\label{jacod}                 
\end{gather}
where we have applied a Fourier transform $g(r,t)\rightarrow g(q,t)$, with $q$ being the wave number. Now trace of the $\mathcal{J_{D}}$ will be simply:
$Tr(\mathcal{J_{D}})=Tr(\mathcal{J})-q^2Tr(\mathcal{D})=k_{2}b-k_{4}-\frac{k_3 {k_1}^2}{{k_4}^2}a^2-(D_{11}+D_{22})q^2$ and determinant of $\mathcal{J_{D}}$ will be
\begin{eqnarray}
 det(\mathcal{J_D})=det(\mathcal{D})q^4-[D_{11}J_{22}+D_{22}J_{11}-D_{12}J_{21} \nonumber \\
 -D_{21}J_{12}]q^2+det(\mathcal{J}),
 \label{deter}
 \end{eqnarray}
a quadratic equation of $q^2$ in which $det(\mathcal{J})=\frac{k_{1}^2 k_{3}}{k_4}a^2$ is determinant of  $\mathcal{J}$. Eigenvalues $\lambda$ of $\mathcal{J_{D}}$ are given by the characteristic equation 
$$\lambda^2-Tr(\mathcal{J_{D}})\lambda+det(\mathcal{J_{D}})=0.$$
Hence eigenvalues can be expressed only in terms of determinant and trace as,
\begin{equation}
    \lambda_{\pm}=\frac{Tr(\mathcal{J_{D}})\pm\sqrt{Tr(\mathcal{J_{D}})^2-4det(\mathcal{J_D})}}{2}.
    \label{eigenvalue}
\end{equation}
Stability criterion simply  demands both of these eigenvalues have to be negative and thus in terms of trace and determinant, this implies $Tr(\mathcal{J_{D}}) <0$ and $det(\mathcal{J_D}) >0$. As chemical concentrations are real quantities,  eigenvalues are complex conjugate pair $\lambda_{\pm}=\lambda_{r}\pm i\lambda_{i}$  at stable steady  state. Since the system was at stable steady state before adding diffusion with $[(D_{11}+D_{22})q^2]>0$ being always true, trace condition of stability, $Tr({\mathcal{J_{D}}})<0$ remains intact even in the presence of the diffusion. So only way to have diffusion driven instability is by breaking the determinant condition of stability in the presence of diffusion. Therefore, $det(\mathcal{J_D})<0$ in the instability regime and at the onset of Turing instability, $det(\mathcal{J_D})=0$. Now from second law of thermodynamics,  $det(\mathcal{D})>0$ is always true and existence of stable steady state in the absence of diffusion demands, $det(\mathcal{J})>0$. So the only way to satisfy $det(\mathcal{J_D})<0$ condition is
\begin{equation}
[D_{11}J_{22}+D_{22}J_{11}]>[D_{12}J_{21}+D_{21}J_{12}].
\label{nct}
\end{equation}
The above condition implies one of the eigenvalues 
crosses zero to become positive and is a necessary
but not sufficient condition to have Turing 
instability in  presence of cross-diffusion. From the necessary condition stated in eq. \eqref{nct}, 
it appears that in the presence of cross-diffusion
so-called local activation and lateral inhibition for 
traditional Turing pattern need not be followed. 
To obtain the necessary and sufficient condition
for having Turing instability induced spatial pattern,
we need to ensure the existence of the real root of quadratic eq. \eqref{deter}, i.e., to satisfy the 
following condition
\begin{equation}
\begin{split}
(D_{11}J_{22}+D_{22}J_{11}-D_{12}J_{21}-D_{21}J_{12})^2 \\
-4det(\mathcal{D})det(\mathcal{J})>0.
    \label{nsct}
\end{split}
\end{equation}
If we assume that by varying the control parameter, 
$b$ the onset of instability is reached, then the 
condition in eq.  \eqref{nsct} simply results in 
following equality,
\begin{equation}
\begin{split}
    (D_{11}J_{22}+D_{22}J_{11}-D_{12}J_{21}-D_{21}J_{12})^2\\
    -4det(\mathcal{D})det(\mathcal{J})=0.
\label{nscte}
\end{split}
\end{equation}
Inserting all the elements of Jacobian, $\mathcal{J}$ and $det(\mathcal{J})$ into eq. \eqref{nscte}, we will find the critical value of the bifurcation parameter in the Brusselator model as
\begin{widetext}
\begin{equation}
    b_{cT}=
    \left(\frac{[D_{11}\frac{k_{1}^2 k_{3}}{{k_4}^2}+D_{21}\frac{k_{1}^2 k_{3}}{{k_4}^2}]a^2+2[det(\mathcal{D})]^\frac{1}{2}[\frac{k_{1}^2 k_{3}}{k_4}]^\frac{1}{2}a+D_{22}k_{4}}{D_{22}k_{2}+D_{12}k_{2}} \right).
    \label{ccpt}
\end{equation}
\end{widetext}
Eigenvalues at the onset of Turing instability now becomes, $$\lambda_{+}=Tr(\mathcal{J_{D}})=k_{2}b_{cT}-k_{4}-\frac{k_3 {k_1}^2}{{k_4}^2}a^2-(D_{11}+D_{22})q_{cT}^2$$
and  $\lambda_{-}=0$. Here $q_{cT}$ is an intrinsic critical wave number and $b_{cT}$ is critical value of the control parameter at the onset of Turing instability. Negative value of the $Tr(\mathcal{J_{D}})$ means that eigenvalue 
$\lambda_{-}=0$ at the Turing instability will govern the whole dynamics of the system.

Necessary and sufficient condition to have Turing instability is that $det(\mathcal{J_D})$ equation must have double roots at the onset of instability, i.e., following two conditions are satisfied simultaneously:
$det(\mathcal{J_D})=0$ and $\frac{d \{det(\mathcal{J_D})\}}{d(q^2)}=0$. This will result in equation of intrinsic critical wave number at the onset of instability,
\begin{equation}
 q_{cT}=\Bigg[\frac{det(\mathcal{J})}{det(\mathcal{D)}}\Bigg]^\frac{1}{4}=\Bigg[\frac{k_{1}^2 k_{3}}{k_4}\frac{a^2}{det(\mathcal{D})} \Bigg]^\frac{1}{4}
    \label{cw}
\end{equation}
and it  will set the length scale as $\frac{2\pi}{q_{cT}}$. This $q_{cT}$ is the fastest growing Fourier mode and for critical value of the control parameter growth rate first becomes zero at this critical wave number.
Now for Turing instability the critical eigenvector, $U_{cT}$  corresponding to eigenvalue $\lambda_{q_{cT}}=0$ is
\begin{gather}
    U_{cT}=
    \begin{pmatrix}
    1\cr
    -\frac{k_4}{(D_{12}+D_{22}){q_{cT}}^2}-\frac{(D_{21}+D_{11})}{(D_{12}+D_{22})}
    \end{pmatrix}\nonumber\\
    =
    \begin{pmatrix}
       1\cr
    -\frac{k_4}{k_1}\sqrt{\frac{k_4}{k_3}}\frac{\sqrt{det(\mathcal{D})}}{(D_{12}+D_{22})a}-\frac{(D_{21}+D_{11})}{(D_{12}+D_{22})}
    \end{pmatrix}.
\end{gather}
Above the critical parameter value, a quite small but finite band of Fourier modes in the vicinity of critical wave number, $q_{cT}$ are considered to be equally excited and thus contribute to nonlinear growth of the spatial pattern. However, in  a finite system with length $l$ subjected to zero flux boundary condition, accessible critical wave number will be given by $q_{cT}=\frac{n\pi}{L}$ for Turing instability. One needs to set the integer value, $n$ in such a way that the admissible critical wave number is nearest to intrinsic critical wave number, $q_{cT}$. 

The circumstances for Turing instability in the Brusselator are more favourable if $D_{21}$ is negative and $D_{12}$ is positive\cite{Nkumar, Zemskov}. However, too much negative $D_{21}$ or positive $D_{12}$  may supress the Turing instability in the Brusselator model\cite{Nkumar}. The conditions in cross diffusion coefficients to obtain favourable circumstances of Turing instability will be inverted in the case of the model like Oregonator because of the opposite cross kinetic behavior compared to the Brusselator model.

\subsection{\label{hop}Hopf instability}

Besides diffusion driven Turing instability, reaction diffusion system could also have a type III-o\citep[ch.\ 10]{Cross2009PatternSystems} oscillatory Hopf instability with critical wave number $q_{cH}=0$. For Hopf instability as the control parameter is varied, trace condition of the stability will be broken as $Tr(\mathcal{J_D})_{|q=0}$ moves to the positive value but initial determinant condition holds. So at the onset of Hopf instability, $Tr(\mathcal{J_D})_{|q=0}=0$, i.e.,  $J_{11}+J_{22}=0$ or $J_{11}=-J_{22}$ and this condition leads to  critical value of control parameter as
\begin{equation}
 b_{cH}=\frac{k_{4}}{k_{2}}+\frac{k_{1}^2 k_{3}}{k_{2}{k_4}^2}a^2.
 \label{hbc}
\end{equation}
Real parts of  complex conjugate eigenvalues which are negative initially will be zero at $b=b_{cH}$ and eigenvalues can be expressed only in terms of determinant of the Jacobian matrix following from eq. \eqref{eigenvalue} as
\begin{equation}
 \lambda_{\pm}=\pm i \sqrt{det(\mathcal{J_D})_{|q=0}}=\pm i \sqrt{\frac{k_{1}^2 k_{3}}{k_4}}a.
 \label{hopfg}
\end{equation}
Critical frequency of Hopf bifurcation, $\omega_{cH}$ is given  by the imaginary part of the eigenvalue at the onset of instability. Therefore, the period of the limit cycle near the the Hopf instability, i.e., slightly above $b_{cH}$ is approximately, $T=\frac{2\pi}{\omega_{cH}}$, where $\omega_{cH}=\sqrt{\frac{k_{1}^2 k_{3}}{k_4}}a$ for the Brusselator model. Critical eigenvector, $U_{cH}$ corresponding to eigenvalue, $\lambda=i{\sqrt{det(\mathcal{J})}}$  at the onset of Hopf instability in the Brusselator model is
\begin{gather}
U_{cH}=
\begin{pmatrix}
=1+i\frac{\sqrt{det(\mathcal{J})}}{J_{11}}\cr
\frac{J_{21}}{J_{11}}
\end{pmatrix}
=
\begin{pmatrix}
1+\frac{i}{a}\sqrt{\frac{k_4}{k_3}}\frac{1}{k_1}\cr
       -(1+\frac{{k_4}^3}{k_3 {k_1}^2}\frac{1}{a^2}) 
\end{pmatrix}.
\end{gather}

\section{\label{sec:amplitudeequation}Derivation of amplitude equation  using Krylov-Bogolyubov method}

Amplitude is a complex entity  often features characteristics analogous to that of the order parameter in phase transition\citep{aranson2002world} and its profile in pattern formation shows a pitchfork bifurcation in a system with translational symmetry.  Turing and Hopf interplay, their relative strength and stability can be studied by exploiting the analytic solutions of their respective amplitude equations.

The Krylov-Bogolyubov(KB) averaging method is a standard method for analysis of oscillation in nonlinear mechanics\citep{krylov1949introduction}. The essential idea of this averaging method consists of varying the magnitude and phase of the amplitude so slowly in time and space that the solution of averaged system approximates the exact dynamics. Introducing two new variables, namely the total concentration of internal species, $z =x-y$, and $u=a-x$, it is possible to rewrite eq. \eqref{dynamic} of the Brusselator model.  Substitution of new variables and with simplification of all the forward rate constants setting as unity, the steady state solution is given by  $u_s=0$ and $z_s=\frac{b}{a}+a$. To shift the fixed point into the origin  a new variable $\zeta=z-z_s$ has been introduced  to obtain, 
a single second order equation  with a form quite similar to that of the generalized Rayleigh equation\citep{Rayleigh1945TheSound} as,
 \begin{equation}
    \ddot{\zeta}+\Omega^2\zeta=\lambda[2(1+c_1u-c_2u^2)u-\frac{1}{\lambda}(u^2-2\Omega u)\zeta]
    \label{eq:3}
\end{equation}
where  $\Omega=a, \lambda=\frac{b-1-a^2}{2}, c_1=\frac{(2a-\frac{b}{a})}{2\lambda}, c_2=\frac{1}{2\lambda}$.
\\Now by taking $2(1+c_1u-c_2u^2)u-\frac{1}{\lambda}(u^2-2\Omega u)\zeta=h$  the eq. \eqref{eq:3} becomes
\begin{equation}
 \ddot{\zeta}+\Omega^2\zeta=\lambda h. 
 \label{zeta}
\end{equation}
Now in the presence of the both self and cross diffusion coefficients which are in general unequal, we can write eq. \eqref{zeta} in the following form 
\begin{eqnarray}
\ddot{\zeta}+\Omega^2\zeta=\lambda h+(D_{22}+D_{12}-D_{11}-D_{21})\dot{u}_{rr}\nonumber\\
+(D_{22}+D_{12})\dot{\zeta}_{rr}
+(D_{11}-D_{12})u_{rr}-D_{12}\zeta_{rr}. \label{dzeta}
\end{eqnarray} 
For very small value of $\lambda$, eq. \eqref{dzeta} admits simple harmonic function like solutions,
\begin{subequations}
\begin{align}
 \zeta(r,t)&=\mathcal{A}(r,t)\cos(\Omega t-\phi(r,t))\label{sh1}\\
 u(r,t)&=\dot{\zeta}(r,t)=-\Omega \mathcal{A}(r,t)\sin(\Omega t-\phi(r,t))\label{sh2}  
\end{align}
\end{subequations}
where both the amplitude, $\mathcal{A}$ and phase, $\phi$ are changing very slowly during fast oscillations. From  eq. \eqref{sh1} and \eqref{sh2}, we can easily find all the required spatial derivatives 
\begin{subequations}
\begin{align}
\zeta_{rr}=(2\mathcal{A}_{r}\phi_{r}+\phi_{rr}\mathcal{A})\sin(\Omega t-\phi)\nonumber \\ 
+(\mathcal{A}_{rr}-\mathcal{A}\phi_{r}^2)\cos(\Omega t-\phi)
\label{szeta}
\end{align}
\begin{align}
u_{rr}=\Omega (\mathcal{A}\phi_{r}^2-\mathcal{A}_{rr})\sin(\Omega t-\phi) \nonumber \\
+\Omega(2\mathcal{A}_{r}\phi_{r}+\phi_{rr}\mathcal{A})\cos(\Omega t-\phi).
\label{su}
\end{align}
\end{subequations}
Further, with the aid of eq. \eqref{sh1} and \eqref{sh2}, we acquire the following form of the amplitude dynamics,
\begin{eqnarray}
\dot{\mathcal{A}}=-\frac{1}{\Omega}[\lambda h-\Omega^2(D_{22}+D_{12}+\frac{D_{12}}{\Omega^2}-D_{11}\nonumber \\
-D_{21})\zeta_{rr}+(D_{22}+D_{11})u_{rr}] \sin(\Omega t-\phi),
\label{amp}
\end{eqnarray}
and  the dynamical equation of phase,
\begin{eqnarray}
\dot{\Phi}=\frac{1}{\Omega \mathcal{A}}[\lambda h-\Omega^2(D_{22}+D_{12}+\frac{D_{12}}{\Omega^2}-D_{11}\nonumber \\
-D_{21})\zeta_{rr}+(D_{22}+D_{11})u_{rr}]\cos(\Omega t-\phi).
\label{phs} 
\end{eqnarray}
Now by taking average over one cycle, fast oscillation parts can be easily ironed out and we obtain amplitude and phase  equations of the Brusselator model in the presence of cross diffusion as
\begin{subequations}
\begin{align}
\dot{\mathcal{A}}=\mathcal{A}\lambda-p_{1}\frac{3\lambda c_{2}\Omega^2}{4}A^3+\frac{\Omega}{2}(D_{22}+D_{12}+\frac{D_{12}}{\Omega^2}-D_{11}\nonumber \\ -D_{21})(2\mathcal{A}_{r}\phi_{r}+\phi_{rr}\mathcal{A})+\frac{(D_{11}+D_{22})}{2}(\mathcal{A}_{rr}-\mathcal{A}\phi_{r}^2), \label{kbamp}\\
\dot{\Phi}=-p_{2}\frac{\Omega}{8}\mathcal{A}^2+\frac{(D_{11}+D_{22})}{2} (\frac{2\mathcal{A}_{r}\phi_{r}}{\mathcal{A}}+\phi_{rr})-\frac{\Omega}{2}(D_{22}\nonumber \\
+D_{12}+\frac{D_{12}}{\Omega^2}-D_{11}-D_{21})(\frac{\mathcal{A}_{rr}}{\mathcal{A}}-\phi_{r}^2).
\label{kbphase}
\end{align}
\end{subequations}
To take into account the effect of non-negative term $2 \lambda c_1$ in eq.\eqref{eq:3} that generates unidirectional acceleration from unstable stationary point,  correction factors, $p_{2}$ and $p_{1} $ in phase shift and limit cycle radius expression needs to be introduced\cite{Lavrova2009PhaseInflux}.

\subsection{\label{hae}Hopf amplitude equation}

The system dynamics near the onset of Hopf instability can be described by using the lowest-order amplitude equation as the complex Ginzburg Landau equation(CGLE)\citep{Kuramoto1984ChemicalTurbulence, aranson2002world, Walgraef1997ThePatterns}. From the phase and amplitude eq.\eqref{kbphase} and \eqref{kbamp} found by KB method, we can arrive at a particular form which agrees with unscaled form of CGLE
\begin{equation}
\frac{\partial Z}{\partial t}=\lambda Z +(\alpha_{r}+ i \alpha_{i})\partial_{r}^2 Z-(\beta_{r}-i\beta_{i})\mid Z \mid ^2 Z.
\label{cgl}
\end{equation}
By setting $Z=\mathcal{A}\exp(-{i \phi})$ in eq. \eqref{cgl} and separating real and imaginary parts one obtains
\begin{subequations}
\begin{align}
\frac{\partial \mathcal{A}}{\partial t}&=\lambda \mathcal{A} -\beta_{r}\mathcal{A}^3+\alpha_{i}(2\mathcal{A}_{r}\phi_{r}+\phi_{rr}\mathcal{A})+\alpha_{r}(\mathcal{A}_{rr}-\mathcal{A}\phi_{r}^2), \label{ampd}\\
\frac{\partial \phi}{\partial t}&=-\beta_{i}\mathcal{A}^2+\alpha_{r}(\frac{2\mathcal{A}_{r}\phi_{r}}{\mathcal{A}}+\phi_{rr})-\alpha_{i}(\frac{\mathcal{A}_{rr}}{\mathcal{A}}-\phi_{r}^2). \label{phased}
\end{align}
\end{subequations}
Eq. \eqref{ampd} and eq. \eqref{phased} are exact deductions of the CGLE and represent amplitude and phase dynamics, respectively near the onset of Hopf instability. Comparing eq. \eqref{ampd} and \eqref{phased} with the dynamical equations of amplitude and phase derived by the KB method, eq. \eqref{kbamp} and \eqref{kbphase}, we obtain all these coefficients, 
$\lambda =\frac{b-1-a^2}{2}$, 
 $\beta_{r} =p_{1}\frac{3\lambda c_{2}\Omega^2}{4}$,  $\beta_{i} =p_{2}\frac{ \Omega}{8}$, $\alpha_{r} =\frac{(D_{11}+D_{22})}{2}$, $\alpha_{i} =\frac{\Omega}{2}(D_{22}+D_{12}+\frac{D_{12}}{\Omega^2}-D_{11}-D_{21})$.
Now, here we will introduce the following scaled variables,

$\mathcal{A}=\frac{\mathcal{A}}{\sqrt{\beta_{r}}}$, $r=\frac{r}{\sqrt{\alpha_{r}}}.$

This scaling will result in the following form of amplitude and phase equations:
\begin{subequations}
\begin{align}
\frac{\partial \mathcal{A}}{\partial t}&=\lambda \mathcal{A} -\mathcal{A}^3+\alpha(2\mathcal{A}_{r}\phi_{r}+\phi_{rr}\mathcal{A}) +(\mathcal{A}_{rr}-\mathcal{A}\phi_{r}^2), \label{samp}\\
\frac{\partial \phi}{\partial t}&=-\beta \mathcal{A}^2+(\frac{2\mathcal{A}_{r}\phi_{r}}{\mathcal{A}}+\phi_{rr})-\alpha(\frac{\mathcal{A}_{rr}}{\mathcal{A}}-\phi_{r}^2), \label{sphase}
\end{align}
\end{subequations}
corresponding to normal form of complex Ginzburg-Landau equation\citep{Nicolis1995IntroductionScience, Walgraef1997ThePatterns,
Cross2009PatternSystems} in one space dimension at the onset of Hopf instability in spatially extended system as
\begin{equation}
   \frac{\partial Z}{\partial t}=\lambda Z +(1+i \alpha)\partial_{r}^2 Z-(1-i\beta)\mid Z \mid ^2Z.
   \label{ncgle}
\end{equation}
Coefficients in normal form of CGLE are now solely the ratio between imaginary and real parts of the complex coefficients of diffusive and nonlinear terms in eq. \eqref{cgl} and is given by
$\alpha=\frac{\alpha_{i}}{\alpha_{r}}=\frac{\Omega(D_{22}+D_{12}+\frac{D_{12}}{\Omega^2}-D_{11}-D_{21})}{(D_{11}+D_{22})}$ and 
$\beta=\frac{\beta_{i}}{\beta_{r}}=\frac{p_2}{p_1}\frac{1}{3a}.$
The coefficient, $\alpha$ found by using KB averaging in the case of the Brusselator model with cross diffusion exactly matches with the one found by using rigorous method of multiscale approach in ref.~\citep{Zemskov2011AmplitudeDiffusion}. It is quite apparent that the coefficient, $\alpha$ depends on the both self and cross diffusion terms explicitly in the case of Hopf instability. Another coefficient, $\beta$  does not have any dependence on diffusion and is given in ref.    \citep{Kuramoto1984ChemicalTurbulence} as $\beta=\frac{4-7a^2+4a^4}{3a(2+a^2)}$ for the Brusselator model. 
Properties of uniform oscillations can be obtained  from eq. \eqref{ncgle} by considering a simple and general state of nonlinear oscillations as 
\begin{equation}
    Z=\mathcal{A}\exp({i\omega_{0}t})
    \label{trialeq}
\end{equation}
where $\omega_{0}$ is the shift in frequency from the critical frequency $\omega_{cH}$ \citep{Cross2009PatternSystems}. Now by inserting it into normal CGLE \eqref{ncgle} and comparing imaginary and real parts, we get $\mathcal{A}^2 =\lambda$ and $\omega_{0} =\beta \mathcal{A}^2=\beta \lambda$. $\beta$ being a nonlinear phase shift, captures dependence  of oscillation frequency on the magnitude of the amplitude and wave number shift. Hence nonlinear oscillations for Hopf bifurcation, can be specified from eq. \eqref{trialeq} as 
\begin{equation}
    \mathcal{A}_{H}=\sqrt{\lambda}\exp({i\beta \lambda t}).
    \label{hopfamp}
\end{equation}

\subsection{\label{ta}Turing amplitude}

Amplitude equation corresponding to Turing instability is known as Turing Amplitude Equation(TAE) which is real counterpart of the  CGLE. Near the onset of Turing instability, lowest order case of one dimensional TAE  can simply be constructed  by symmetry 
argument\citep{Cross2009PatternSystems} as 
\begin{equation}
  \frac{\partial Z}{\partial t}=\lambda_T  Z+\varrho\partial_{r}^2 Z-g\mid Z \mid ^2Z
  \label{tae}
\end{equation}
where $\lambda_T=\frac{b-b_c}{2}$ defined in similar way as $\lambda$ in Hopf instability and $\varrho$ and $g$ are coefficients containing details of the system. By setting $Z=\mathcal{A}\exp(-{i \phi})$ in eq. \eqref{tae} and then separating real and imaginary parts we obtain
\begin{subequations}
\begin{align}
\frac{\partial \mathcal{A}}{\partial t}&=\lambda_T \mathcal{A}- g\mathcal{A}^3 + \varrho(\mathcal{A}_{rr}- \mathcal{A}\phi_{r}^2) \label{ate}\\
\frac{\partial \phi}{\partial t}&=\varrho(\frac{2\mathcal{A}_{r}\phi_{r}}{\mathcal{A}}+\phi_{rr}\label{pte}).
\end{align}
\end{subequations}
We are only  interested in bifurcation scenario of generic dynamical features of the system here and hence it is enough to have only parameter $\lambda_T$ in TAE. Now, by introducing the following scales in amplitude and spatial dimension, $$\mathcal{A}=\frac{\mathcal{A}}{\sqrt{g}},r=\frac{r}{\sqrt{\varrho}}$$ and taking constant phase value by virtue of translational  invariance of spatial pattern in eq.\eqref{ate} and \eqref{pte} we find
\begin{subequations}
\begin{align}
\frac{\partial \mathcal{A}}{\partial t}&=\lambda_T \mathcal{A} -\mathcal{A}^3 + \mathcal{A}_{rr}, \label{fate}\\
\frac{\partial \phi}{\partial t}&=0\label{fpte}
\end{align}
\end{subequations}
where $b_{cT}$ in $\lambda_{T}=\frac{b-b_{cT}}{2}$  is given by the eq. \eqref{ccpt} and it contains all the effect of self-diffusion as well as cross-diffusion constant. Eq. \eqref{fate} and \eqref{fpte} are deduction of normalized form of TAE for the case of constant phase as
\begin{equation}
  \frac{\partial Z}{\partial t}=\lambda_T  Z+\partial_{r}^2 Z-\mid Z \mid ^2Z
  \label{ntae}
\end{equation}
which could be simply regarded as special case of the normal form of CGLE \eqref{ncgle} if $\alpha$ and $\beta$ are set to zero\citep{aranson2002world}. Eq. \eqref{fate} admits time-dependent homogeneous solution of Turing amplitude as  
\begin{equation}
 \mathcal{A_T}^2=\mathcal{A}_{s}^2 \Big[\frac{1}{1-\mathcal{A}_{0}\exp{(-2\lambda_T (t-t_{0}))}}\Big] \label{Turingamp}
 \end{equation}
 which renders $\mathcal{A_T}=\sqrt{\lambda_T}$ for long time limit.

\section{\label{sec:entropy}Entropy production rate}

For a chemical reaction network, fluxes are not a linear function of the conjugate force. Net reaction currents of reversible chemical reactions are given as the difference between forward and reverse fluxes of reactions:
\begin{equation}
    j_{\rho}=j_{+\rho}-j_{-\rho}\label{rcur}
\end{equation}
where $'+'$ and $'-'$ label forward and backward reaction, respectively. Since $k_{-\rho}\simeq 0$ is assumed in sec. \ref{sec:level2}, all the reverse reaction fluxes are negligibly small, i.e., $j_{-\rho}\simeq 0.$ Concentration fluxes according to the law of mass action are,
\begin{equation}
  j_{\pm \rho}=k_{\pm\rho}\prod_{\sigma}z^{v_{\pm\rho}^{\sigma}}_{\sigma} \label{marc} 
\end{equation}
with $v_{\pm \rho}^{\sigma}$ denotes the number of molecules of a particular species $'\sigma'$ for forward(+) or reverse(-) direction of reaction $'\rho'$. Whereas, according to Fick's diffusion law the diffusion current is proportional to the gradient of the concentration distribution of diffusing species and in one dimensional system, simply reduces to
\begin{equation}
J_{\sigma}=-D \frac{\partial z_{\sigma}}{\partial r} 
\label{fick}
\end{equation}
 with constant diffusion coefficient $D$ being one of the  the elements of the matrix $\mathcal{D}=\begin{pmatrix}
 D_{11} & D_{12}\\
 D_{21} & D_{22}
 \end{pmatrix}$ in presence of the cross diffusion.

Product of the stoichiometric coefficient of species $'\sigma'$ of a particular reaction step $'\rho'$ and corresponding chemical potential, $\mu_{\sigma}$ gives thermodynamic driving forces of reaction known as reaction affinities\citep{Prigogine1954ChemicalDefay.}:
\begin{equation}
f_{\rho}=-\sum_{\sigma}{S_{\rho}^{\sigma}\mu_{\sigma}} \label{aff} 
\end{equation} 
where $S_{\rho}^{\sigma}=v_{{-}\rho}^{\sigma}-v_{{+}\rho}^{\sigma}$ \text{and}  $\mu_{\sigma}=\mu_{\sigma}^o+\ln{\frac{z_{\sigma}}{z_0}}$ with solvent concentration $z_0$ and standard-state chemical potential $\mu_{\sigma}^o$. To define a base-line for substances, standard-state quantities with notation $'o'$ are defined  at standard pressure $p=p^{o}$ and molecular concentration and chemical potential, $\mu_{\sigma}$ characterises each chemical species of the dilute solution thermodynamically. System is maintained at constant absolute temperature  $T$ fixed by the solvent, and for simplicity, $RT$ is taken as unity. Using this form of the chemical potential local detailed balance condition of the the reaction steps can be expressed as:
\begin{equation}
\ln{\frac{k_{+\rho}}{k_{-\rho}}}=-\sum_{\sigma}S_{\rho}^{\sigma}\mu_{\sigma}^0. \label{dt}
\end{equation}
Hence reaction affinities in eq. \eqref{aff} can be written in terms of the reaction fluxes of the chemical steps as
\begin{equation}
 f_{\rho}= \ln{\frac{j_{+\rho}}{j_{-\rho}}}.
 \label{rff}
\end{equation}
Eq. \eqref{dt} is very important one as it relates dynamical term with thermodynamic entity. Similar to reaction affinity a thermodynamic driving force, local diffusion affinity exists in the reaction-diffusion system and can be expressed as a gradient of the chemical potential
\begin{equation}
    F_{\sigma}=-\frac{\partial \mu_{\sigma}}{\partial r}  \label{force}.
\end{equation}
Entropy production rate(EPR) due to the chemical reaction can be expressed as the product of the thermodynamic driving force and reaction flux as
\begin{equation}
    \frac{d\Sigma_{R}}{dt}=\frac{1}{T}\int  dr \sum_{\rho}{f_{\rho}j_{\rho}}\label{epr}.
\end{equation}
So with the help of  the eq. \eqref{rff} and eq. \eqref{rcur}, we obtain EPR due to reaction as
\begin{equation}
  \frac{d\Sigma_{R}}{dt}=\frac{1}{T}\int  dr \sum_{\rho} (j_{+\rho}-j_{-\rho}) \ln{\frac{j_{+\rho}}{j_{-\rho}}},
  \label{eprr}
\end{equation}
considering the elementary reaction steps which are  directly related to reaction stoichiometry.

Similarly, entropy production rate due to diffusion can be simply
\begin{equation}
    \frac{d\Sigma_{D}}{dt}=\frac{1}{T}\int dr \sum_{\sigma}{F_{\sigma}J_{\sigma}}.\label{eprd}
\end{equation}
Considering diffusive flux and affinity given in the eq. \eqref{fick} and \eqref{force}, respectively, we have 
\begin{eqnarray}
 \frac{d\Sigma_{D}}{dt}=\int dr \Big[ D_{11}\frac{{\parallel{\frac{\partial x}{\partial r} }\parallel}^2}{x}+D_{22}\frac{{\parallel{\frac{\partial y}{\partial r} }\parallel}^2}{y}\nonumber\\ +D_{12}\frac{{\parallel{\frac{\partial y}{\partial r}\parallel} {\parallel\frac{\partial x}{\partial r} }\parallel}}{x}+
D_{21}\frac{{\parallel{\frac{\partial x}{\partial r}\parallel} {\parallel\frac{\partial x}{\partial r} }\parallel}}{y}\Big] \label{eprdd}.
\end{eqnarray}
The last two terms on the right side in eq. \eqref{eprdd} correspond to the cross diffusion coefficients of the intermediate species present in the system.

Total entropy production is simply the sum of reaction EPR and diffusion EPR,
\begin{equation}
   \dot{\Sigma}= \dot{\Sigma}_{R}+\dot{\Sigma}_{D} \geq 0 .
\label{tepr}
\end{equation}

In contrast, a closed system must relax to the thermodynamic equilibrium and as a consequence concentration distribution of the species will be distributed homogeneously over the system. At thermodynamic equilibrium all the internal fluxes, i.e., $j_{\rho}$ and $J_{\sigma}$ and external fluxes of the chemostatted species vanish and so the total entropy production rate is zero.

\section{\label{sec:gibbs}Nonequilibrium Gibbs free energy of chemostatic system}

Nonequilibrium Gibbs free energy of a chemical reaction network can be expressed in terms of the Gibbs free energy of an ideal dilute solution\citep[ch.\ 7]{Fermi1956Thermodynamics} as 
\begin{equation}
G=G_{0}+\int dr \sum_{\sigma \neq 0}{(z_{\sigma}\mu_{\sigma}-z_{\sigma})}\label{neg}
\end{equation}
where $G_{0}=z_{0}\mu_{0}^o$. Constant term like $\ln{z_{0}}$ is also absorbed within $\mu_{\sigma}^o$ term of chemical potential. Here solvent has been treated as a special chemostatted element and both $G^{0}$ and $\sum_{\sigma \neq 0} z_{\sigma}$ in eq. \eqref{neg} are due to solvent of dilute solution. For a closed system, when concentration distribution is relaxed to a unique equilibrium distribution, $z_{\sigma}^{eq}$, from eq. \eqref{neg} one finds
\begin{equation}
 G(z_{\sigma}^{eq})=G_{0}+\int dr \sum_{\sigma \neq 0}{(z_{\sigma}^{eq} \mu_{\sigma}^{eq}-z_{\sigma}^{eq})}.\label{eg} 
\end{equation}
Left null vectors corresponding to left null space of the  stoichiometric matrix are  known as the conservation laws\citep[p.89-103]{Alberty2003ThermodynamicsReactions}, whereas the (right) null eigenvectors of the stoichiometric matrix represent cycles. So mathematically, conservation law can be expressed as 
 \begin{equation}
 \sum_{\sigma}{l_{\sigma}^{\lambda}S_{\rho}^{\sigma}}=0 \label{cons}
 \end{equation} 
 where $$S_{\rho}^{\sigma}\in \mathbb{R}^{\sigma \times \rho},\{l_{\sigma}^{\lambda}\}\in \mathbb{R}^{(\sigma-w )\times \sigma}, w=rank(S_{\rho}^{\sigma}).$$
From the definition of affinity in eq. \eqref{aff}, we can further express chemical potentials in terms of linear combination of conservation laws for closed system at equilibrium,
\begin{equation}
 \mu_{\sigma}^{eq}=\mathit{R_{\lambda}}l_{\sigma}^{\lambda}
\label{eqcp}
\end{equation}
where $\mathit{R_{\lambda}}$ is real coefficient with dimension of force. Conserved quantities of closed system known as components can be specified in terms of this conservation laws of the reaction network with 
\begin{equation}
    L_{\lambda}=\sum_{\sigma}{l_{\sigma}^{\lambda}}z_{\sigma}
\end{equation}
such that $\frac{d}{dt}\int dr L_{\lambda}=0$.
Since components $L_{\lambda}$ remain constant over time for closed  reaction diffusion system, it would characterize both the equilibrium and nonequilibrium concentration distribution as $\mu_{\sigma}^{eq}z_{\sigma}^{eq}=\mu_{\sigma}^{eq}z_{\sigma}$. This renders another form of eq. \eqref{eg} as
\begin{equation}
 G(z_{\sigma}^{eq})=G_{0}+\int dr \sum_{\sigma \neq 0}{(z_{\sigma} \mu_{\sigma}^{eq}-z_{\sigma}^{eq})}.\label{ega}   
\end{equation}
Using the relation in eq. \eqref{eqcp} we could also have the  following expression of equilibrium Gibbs free energy from  eq. \eqref{eg}
\begin{equation}
 G^{eq}=G(z_{\sigma}^{eq})=G_{0}+\int dr \sum_{\sigma \neq 0}{(\mathit{R_{\lambda}}L_{\lambda}-z_{\sigma}^{eq})}.
 \label{egm}    
\end{equation}
In information theory approach\citep{Cover1999ElementsTheory}, Shannon entropy or Kullback-Leibler divergence is defined for two normalised probability distributions, $P$ and $P^{0}$ as $$\Gamma(P||P^{0})=\sum_{i}{P_{i}\log \frac{P_{i}}{{P_i}^0}}$$ and it quantifies the amount of information needed to switch from a known distribution $P^{0}$ to the distribution  $P$. With the similar spirit, we can express nonequilibrium Gibbs free energy by exploiting equations \eqref{neg} and \eqref{ega} as
\begin{equation}
    G-G^{eq}=\Gamma(z_{\sigma}||z_{\sigma}^{eq})
    \label{se}
\end{equation}
where 
$$\Gamma(z_{\sigma}||z_{\sigma}^{eq})=\sum_{\sigma}\left\{{z_{\sigma}\log \frac{z_{\sigma}}{z_{\sigma}^{eq}}}-(z_{\sigma}-z_{\sigma}^{eq})\right\}\geq 0,$$ is relative entropy for non-normalized concentration distribution. Thus eq. \eqref{se} implies that the lowest possible value of nonequilibrium Gibbs free energy is set by its equilibrium counterpart in closed system. 

Conservation laws in open system could be characterized in general by
\begin{equation}
l_{I}^{\lambda}S_{\rho}^{I}+l_{C}^{\lambda}S_{\rho}^{C}=0
\begin{cases}
            l_{I}^{\lambda_{b}}S_{\rho}^{I}\neq 0 &\text{ broken CL},\\
            l_{I}^{\lambda_{u}}S_{\rho}^{I}= 0&\text{ unbroken CL}\label{condi}
\end{cases}
\end{equation}
where for open system, $\{l^{\lambda}\}=\{l^{\lambda_b}\}\cup\{l^{\lambda_u}\}$, labels $u$ and $b$ correspond to unbroken and broken ones, respectively. So from eq. \eqref{condi}, we can say broken conservation laws are not left null vectors of $S_{\rho}^{I}$ for at least one reaction of the reaction network. Consequently, corresponding broken components, $L_{\lambda_b}$ of open system are no longer a global conserved quantities. Depending on whether chemostatted species break a conservation law or not, set of chemostatted species could  thus be divided into two subsets so that $\{C\}=\{C_{b}\}\cup\{C_{u}\}$.

The semigrand Gibbs free energy of the open system can be acquired from the nonequilibrium Gibbs free energy as,
\begin{equation}
 \mathcal{G}=G-\sum_{C_b}{\mu_{C_b}^{eq}M_{C_b}}. 
 \label{smgg}
\end{equation}
 where $M_{C_b}=\sum_{C_b}l_{C_b}^{{\lambda_b}^{-1}}\int dr L_{\lambda_b}$ resembles moieties that exchanged between chemostats and system only through the external flow of the chemostatted species. For open system nonequilibrium semigrand Gibbs free energy will  have the form similar to its nonequilibrium Gibbs free energy counterpart,
 \begin{equation}
    \mathcal{G}=\mathcal{G}^{eq}+\Gamma(z_{\sigma}||z_{\sigma}^{eq}).
    \label{osgg}
 \end{equation}
At local level, exchange of  species between neighboring spaces due to local diffusion will be equivalent to matter exchange through chemostating. So energetic contribution of the species exchanged through the local diffusion needs to be eliminated to define proper thermodynamic potential  at local level of the open  reaction diffusion system\citep{Avanzini2019ThermodynamicsWaves}. So from local standpoint, the transformed Gibbs free energy would have the following form,
 \begin{equation}
 \mathcal{G_L}=G-\mu_{\sigma}^{eq}z_{\sigma}
 \label{lgf}
 \end{equation}
where $G$ is Gibbs free energy of the closed system specified at each point of the system. When all the conservation laws are broken then $\mu_{I}^{eq}z_{\sigma}=\mu_{C_b}^{eq}l_{C_b}^{{\lambda_b}^{-1}}l_{I}^{\lambda_b}z^I$ and $\mu_{C_b}^{eq}Z_{\sigma}=\mu_{C_b}^{eq}l_{C_b}^{{\lambda_b}^{-1}}l_{C_b}^{\lambda_b}z^{C_b}$ and thus eq. \eqref{lgf} would result in expression identical to eq. \eqref{smgg}.

For the stoichiometric matrix \eqref{st} of the Brusselator reaction network, the conservation laws of the closed  reaction diffusion system are represented by two linearly independent $(1\times 6)$ vectors,
  \begin{gather}
l_{\sigma}^{\lambda=1}=  
\bordermatrix{~&X&Y&A&B&D&E\cr
             &1&1&1&0&0&1\cr}\label{ce1}
             \end{gather}
             and
  \begin{gather}  
 l _{\sigma}^{\lambda=2}=   
  \bordermatrix{~&X&Y&A&B&D&E\cr
             &0&0&0&1&1&0\cr}.\label{ce2}
                  \end{gather}
The components corresponding to these two conservation laws are $L_1=x+y+a+e$ and $L_2=b+d$. The species $A$ and $B$ are considered as the reference chemostatted species here and both the conservation laws of the Brusselator model in equations \eqref{ce1} and \eqref{ce2} are broken by chemostatting of $A$ and $B$.

\section{\label{confld}Concentration fields of intermediate species}

As mentioned earlier in sec. \ref{hop}, we can have both Turing and Hopf instabilities in  reaction diffusion system. Spatiotemporal profile of the concentration fields in different range of control parameter shows periodic behavior depending on the dispositions of Turing and Hopf instabilities. The  resulting pattern can be traced in the  critical wavenumbers and frequencies of Turing and Hopf regimes from the solution of the corresponding  amplitude equations.

\subsection{\label{turingcon}Turing instability regime}

For the marginal stability condition of the homogeneous state of the system, growth rate becomes zero and the evolution equation of the concentration field near the onset of Turing instability can be expressed by using amplitude equation formalism for the single fastest-growing mode as
\begin{equation}
{z_I}_{T}={z_I}_{0}+A_{T}U_{cT}\exp{(i q_{cT}r)}+C.C
\label{twave}
\end{equation}
where ${Z_I}_{0}\in [x_0,y_0]$ is time-independent uniform base state with respect to extended direction and $A_{T}$ is Turing amplitude rendering several essential features of the pattern formation. The corresponding long time solution of eq. \eqref{twave} is given by
\begin{gather}
\begin{pmatrix}
x\cr
y
\end{pmatrix}=
\begin{pmatrix}
x_0\cr
y_0
\end{pmatrix}
+ 
\bigg[
\begin{pmatrix}
       1\cr
    -\frac{k_4}{k_1}\sqrt{\frac{k_4}{k_3}}\frac{\sqrt{det(\mathcal{D})}}{(D_{12}+D_{22})a}-\frac{(D_{21}+D_{11})}{(D_{12}+D_{22})}
    \end{pmatrix}\nonumber \\
    \times
    A_{T}2\cos{q_{cT}r}\bigg].
    \label{ftwave}
\end{gather}

\subsection{\label{hopfcon}Hopf instability regime}

Similar to the Turing instability in subsec. \ref{turingcon}, perturbation part in the Hopf instability can be written as 
\begin{equation}
\delta{z_I}=A_{H}U_{cH}\exp{(i\omega_{cH}t)}+ C.C
\label{hwave}
\end{equation}
where $A_{H}$ is Hopf amplitude part. Final equation of the perturbation can be expressed as 
\begin{widetext}
\begin{gather}
  \begin{pmatrix}
  \delta{x} \cr
   \delta{y}
  \end{pmatrix}
   = \sqrt{\lambda}\exp({i\beta \lambda t})
   \begin{pmatrix}
2\cos{(\omega_{cH}t+\beta \lambda t)}-\frac{2}{a}\sqrt{\frac{k_4}{k_3}}\frac{1}{k_1}\sin{(\omega_{cH}t+\beta \lambda t)}\cr
       -2(1+\frac{{k_4}^3}{k_3 {k_1}^2}\frac{1}{a^2}\cos{(\omega_{cH}t+\beta \lambda t)})
\end{pmatrix}.
\end{gather}
\end{widetext}
For the parameter value greater than the critical value $b_{cH}$, this perturbation part will give rise to the limit cycle type oscillatory profile.

\subsection{\label{thinteraction}Overlapping of Turing and Hopf instabilities}

Here we are going beyond Turing condition of pattern formation and we have chosen equal self diffusion coefficients and non-zero cross diffusion coefficients  for the  Brusselator model. Critical values of the control parameter, $b$ for Turing and Hopf instabilities are given previously by eq. \eqref{ccpt} in sec. \ref{thins} and \eqref{hbc} in sec. \ref{hop}, respectively. Equating these we can simply derive a particular point, $a_{TH}$ in the parameter space of $a$ for which thresholds of Turing and Hopf instabilities would  coincide in  $(a,b)$  parameter plane as
\begin{widetext}
\begin{equation}
    a_{TH}=\left[\frac{k_4^3}{k_3 k_1^2}\right]^{\frac{1}{2}} \left(\frac{[det(\mathcal{D})]^\frac{1}{2}+\sqrt{[det(\mathcal{D})]-[D_{22}+D_{12}-D_{11}-D_{21}]D_{12}}}{[D_{22}+D_{12}-D_{11}-D_{21}]}\right).
    \label{ath}
\end{equation}
\end{widetext}
In the vicinity of Turing-Hopf point, critical intrinsic wave number of Turing instability obtained from marginal stability condition is given by $$q_{cT}|_{a=a_{TH}}=q_{cTH}=\Bigg[\frac{k_{1}^2 k_{3}}{k_4}\frac{a_{TH}^2}{det(\mathcal{D})} \Bigg]^\frac{1}{4}$$ 
and critical frequency of a homogeneous Hopf mode is $\omega_{cTH}=a_{TH}$. Superposition of Turing mode in eq. \eqref{twave} and Hopf mode in eq. \eqref{hwave} will describe the  spatio-temporal dynamics of the concentration field due to Turing-Hopf interplay with
 \begin{equation}
 \begin{split}
{z_I}_{TH}={z_I}_{0}+A_{T}U_{cT}\exp{(i q_{cT}r)}+A_{H}U_{cH}\exp{(i \omega_{cH}t)}\\ +C.C. \label{wave}
\end{split}
 \end{equation}
This concentration field is employed to assess all the thermodynamic entities corresponding to Turing-Hopf interplay.

We have considered spatio-temporal pattern arising from the interplay between  Turing and Hopf instabilities  in either of following three ways:

\begin{itemize}
\item Stationary spatial Turing pattern grows before it loses stability as control parameter, b is further changed and Hopf instability appears in reaction-diffusion system.\item Homogeneous oscillatory pattern emerges first and then limit cycle solution modulated by Turing instability.\item Critical points of  Turing and Hopf instabilities overlap and thus they arise simultaneously in the system and interact. 
\end{itemize}

Based on amplitude equation formalism in the presence of cross-diffusion, we obtain concentration profiles for all three scenarios with the aid of eq. \eqref{ath} in the space of control parameter, b which lies in the vicinity of onset of instabilities.

\section{\label{RD}Results and discussions}

The evolution of entropy production rate and semigrand Gibbs free energy in the 1D Brusselator model regarded as open chemical network, have been investigated analytically to find out the correspondence between the evolution of thermodynamic quantities and spatiotemporal pattern due to Turing-Hopf interplay. All the results correspond to a steady state condition with absolute temperature: $T=300K$, diffusion coefficients: $D_{11}=D_{22}=1; D_{12}=0.51; D_{21}=-0.51$, one dimensional system length, $l=9.5$ and for weakly reversible case, i.e., chemical reaction rate constants $k_{-\rho}=10 ^{-4} << k_{\rho}=1$ unless otherwise indicated. The temperature is constant throughout the system as rate of heat diffusion is assumed to be much faster than the diffusion rate of species. We have used $b$ as control parameter to find out its effect on the intermediate species concentrations and thus on the thermodynamic entities also.

\begin{figure*}[tb!]
\centering
\subfigure[\label{parameter2}]{\includegraphics[width=0.45\textwidth]{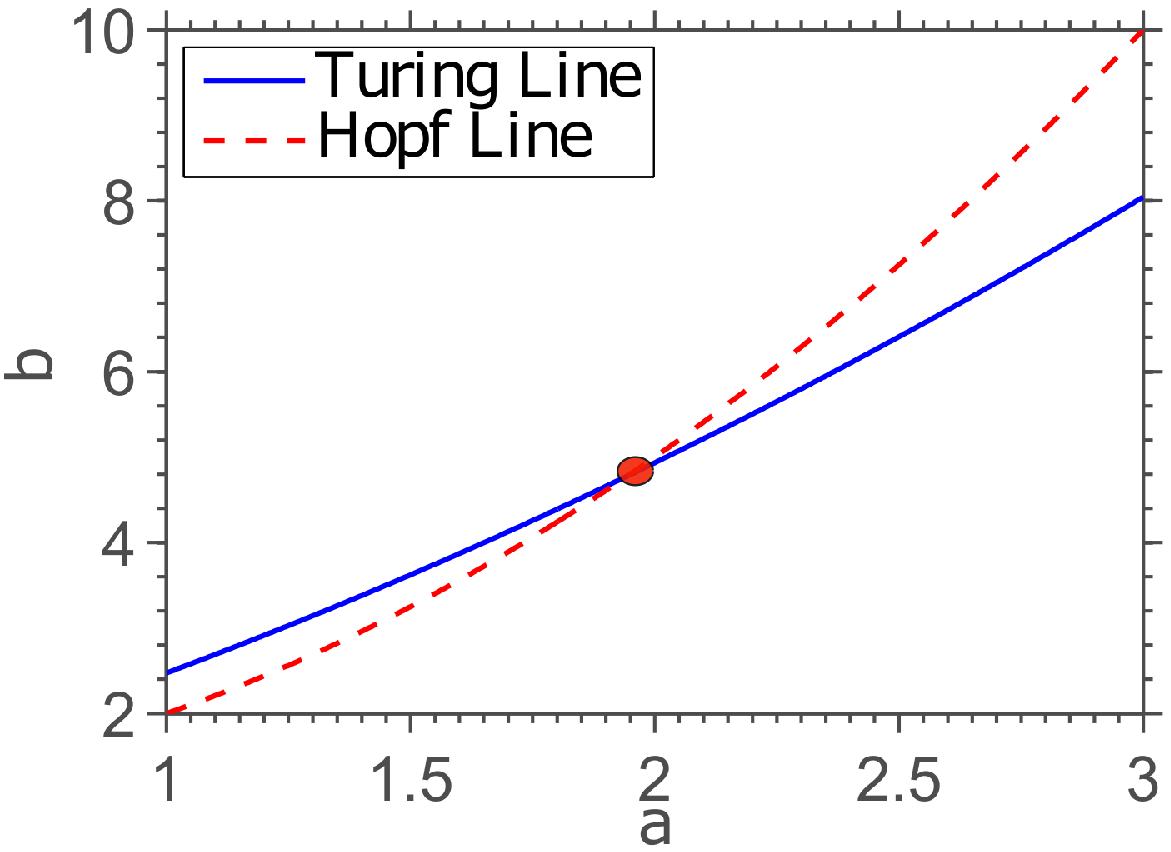}}\hfill
\subfigure[\label{parameter}]{\includegraphics[width=0.45\textwidth]{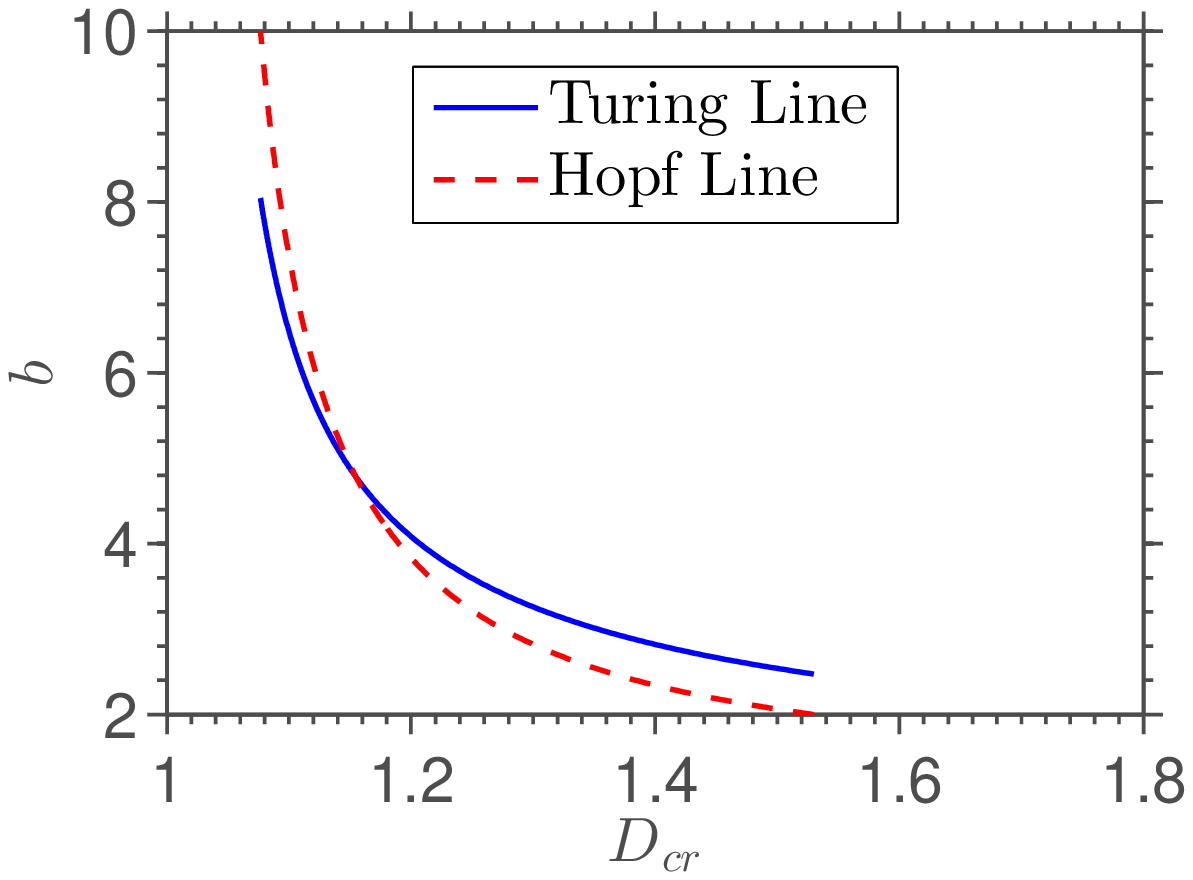}}
\subfigure[\label{parameter2exp}]{\includegraphics[width=0.45\textwidth]{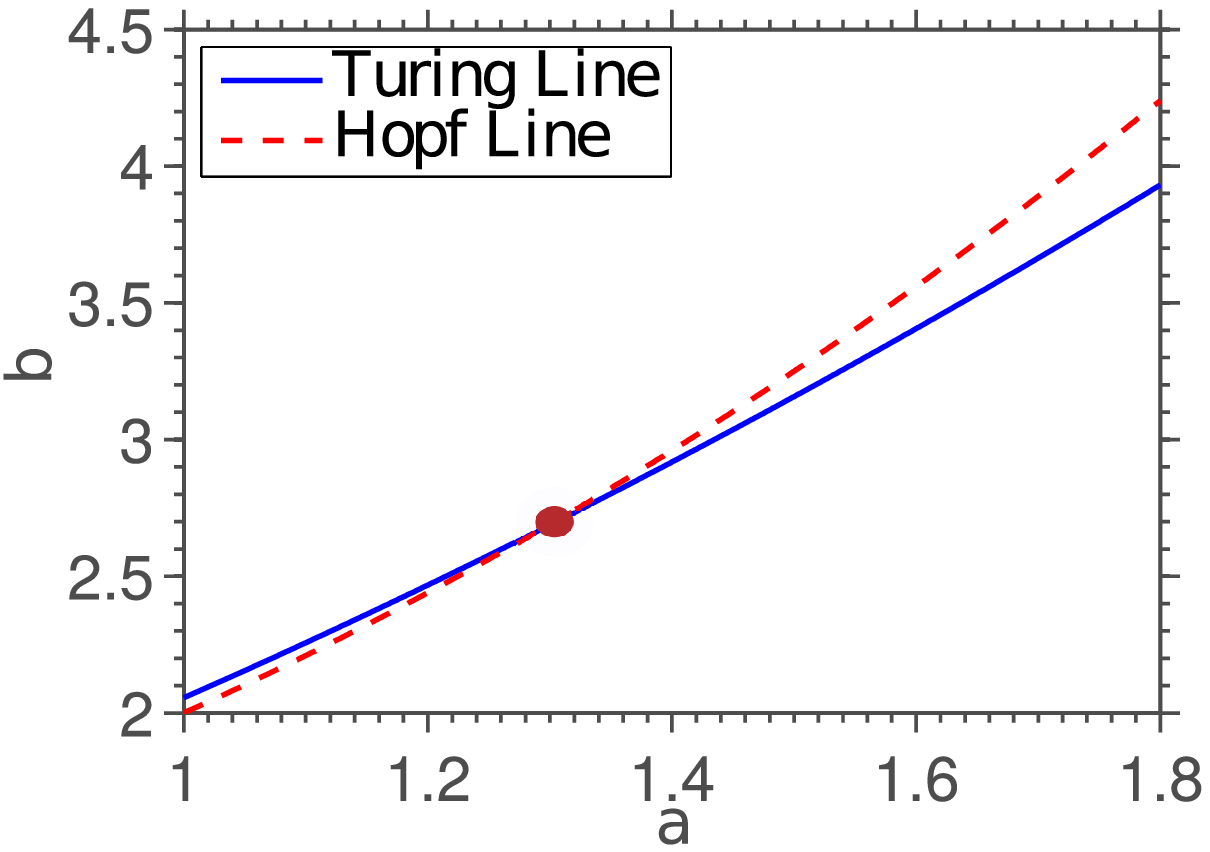}}\hfill
\subfigure[\label{parameterexp}]{\includegraphics[width=0.45\textwidth]{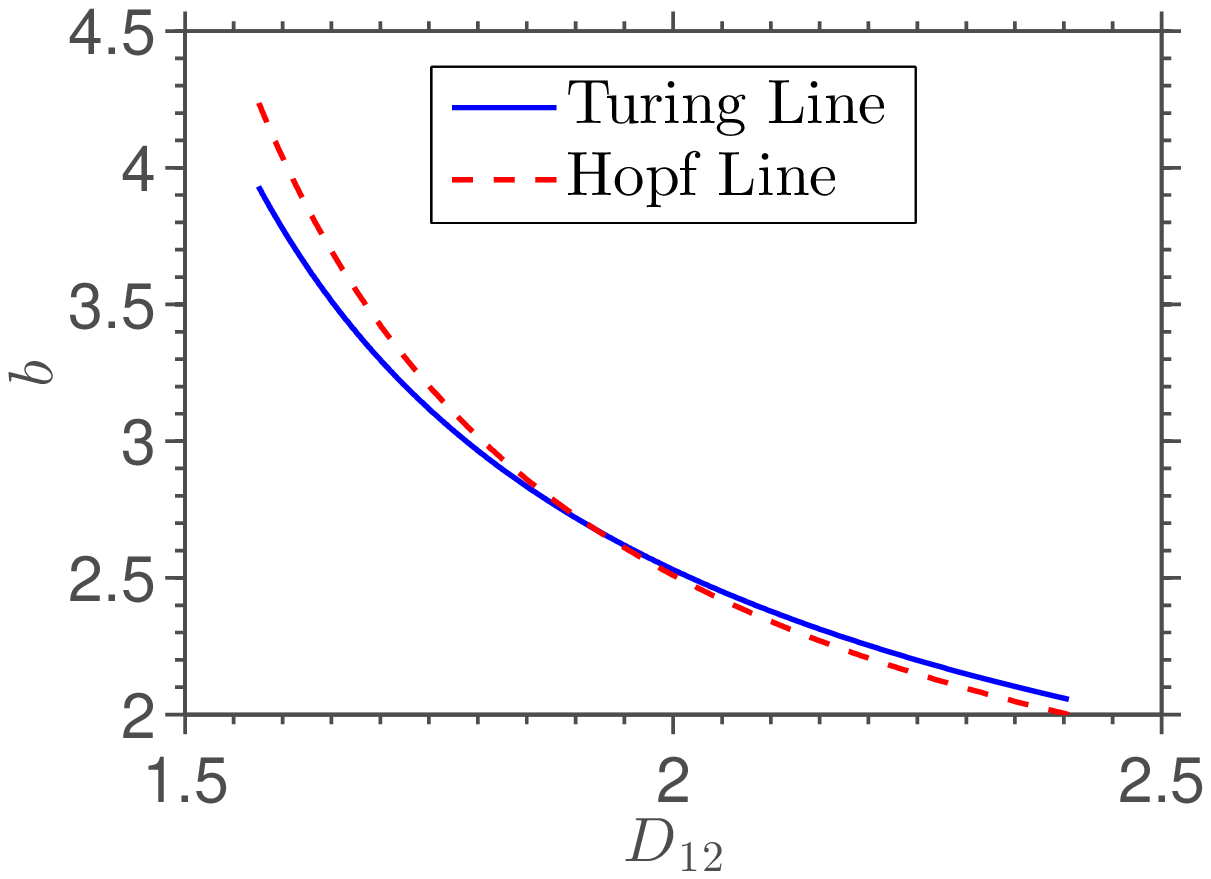}}
\caption{The solid line is Turing line corresponding to eq. \eqref{ccpt}  and dashed curve is Hopf line corresponding to eq. \eqref{hbc} in both FIG. \ref{parameter2} and \ref{parameter} in the presence of diffusion coefficients: $D_{11}=D_{22}=1;D_{12}=0.51;D_{21}=-0.51$. In FIG. \ref{parameter2}, the point of intersection of Turing and Hopf line is [$a_{TH}\approx  1.9438$, $b_{cT}=b_{cH} \approx 4.7785$]. In FIG. \ref{parameter}, we have defined a new parameter, $D_{cr}=D_{12}[1+\frac{1}{a^2}]-D_{21}$ containing only cross diffusion coefficients to explore the effect of cross diffusion on Turing and Hopf line. FIG. \ref{parameter2exp} and \ref{parameterexp} are similar to the previous figures with experimental magnitudes of the diffusion coefficients as $D_{11}=1.28; D_{12}=1.26; D_{21}=-0.005; D_{22}=1.51$ of the pentanary $BZ-AOT$ system.}
\label{fig:parameterspace}
\end{figure*}

In FIG. \ref{parameter2}, we show the region of Turing-Hopf interplay in $(b,a)$ parameter space. We have obtained  Turing line by using eq. \eqref{ccpt} and Hopf line by using eq. \eqref{hbc}  as shown by solid line and dashed line, respectively. The circular label in the figure corresponds to critical Turing-Hopf point(CTHP) where Turing and Hopf line intersect. 

Exploiting the modified Taylor dispersion method, the values of the self diffusion, as well as cross-diffusion coefficients, are experimentally determined in the case of three-component \cite{vanagcross2}, four-component\cite{vanagcross2, Rossi4} and five-component \cite{Rossi5} $BZ-AOT$ system\cite{BZAOT1, BZAOT2}. In their work, Rossi et al. reported that experimentally found cross diffusion coefficients can shift the Turing onset and thus can generate Turing pattern if the system  was initially close to the onset of instability. To obtain proper insight of this experiment claim, we have also taken diffusion matrix elements from the experimental data of pentanary $BZ-AOT$ system\cite{Rossi5} as $D_{11}=1.28; D_{12}=1.26; D_{21}=-0.005; D_{22}=1.51$ on the ground of the assumption that the presence of additional components in the system leaves the diffusion coefficients unchanged\cite{Rossi4} and corresponding results are shown in FIG. \ref{parameter2exp} and \ref{parameterexp}. One should note that the effect of cross diffusion on the onset of Turing instability for the $BZ-AOT$ system was reported in the presence of two different self diffusion coefficients. Although from FIG. \ref{parameter}, it is clear that $D_{cr}$ comprising of cross diffusion coefficients can control the onset of instabilities even when all the self diffusion coefficients are equal\cite{Zemskov2011AmplitudeDiffusion}. 
\begin{figure*}[tb!]
\centering
\subfigure[\label{parameter2_cd}]{\includegraphics[width=0.45\textwidth]{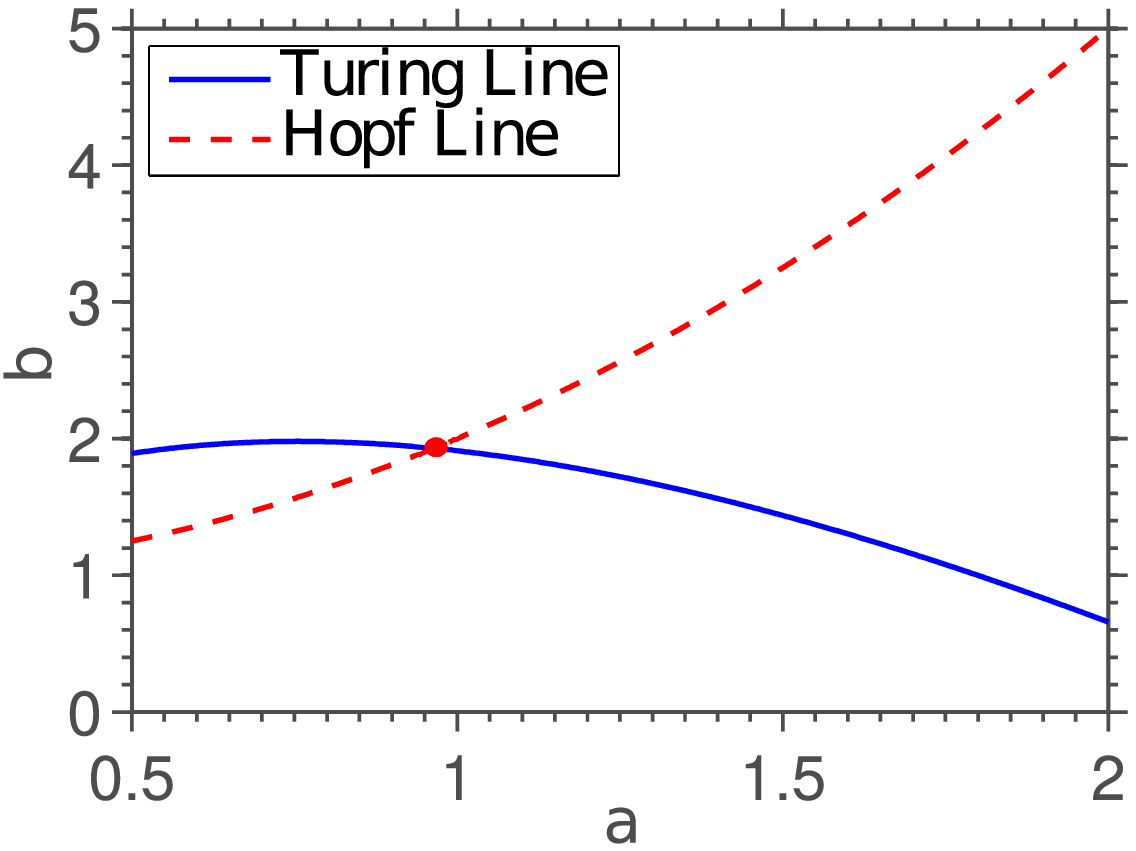}}\hfill
\subfigure[\label{parameter_cd}]{\includegraphics[width=0.45\textwidth]{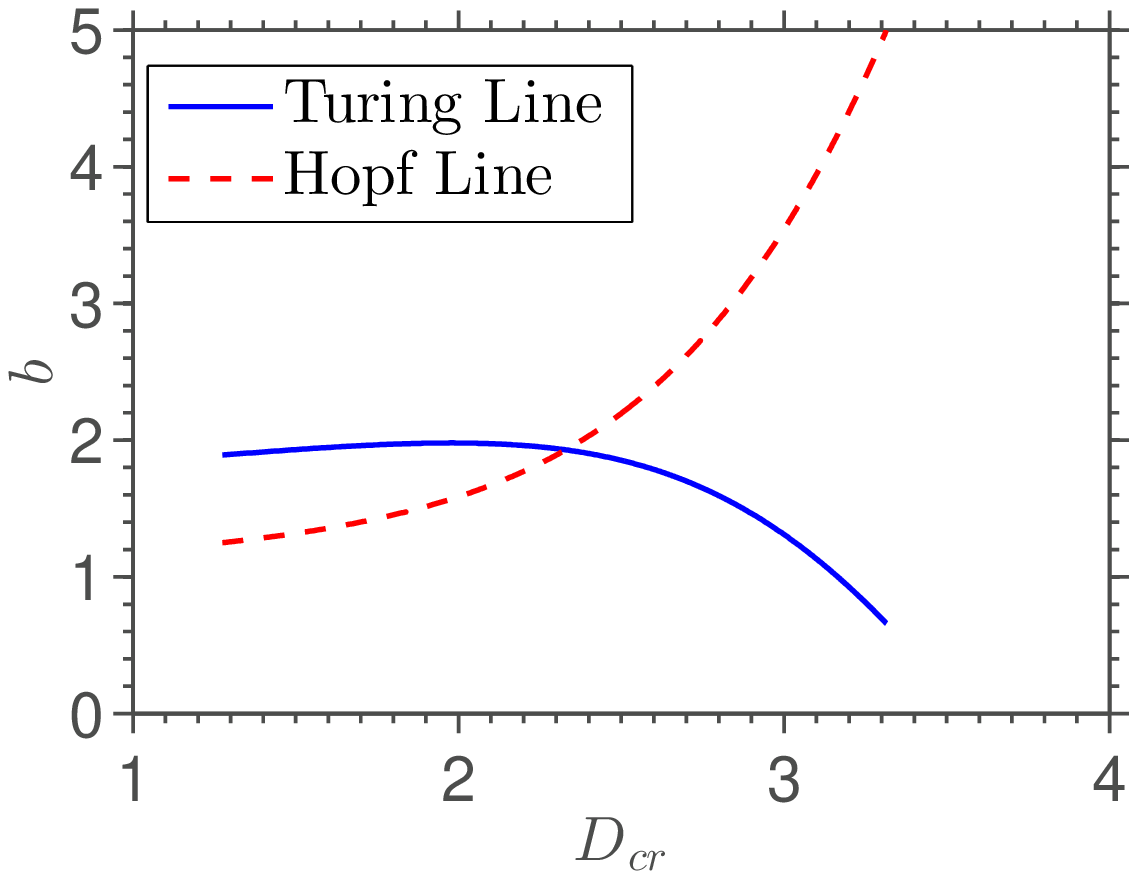}}
\subfigure[\label{b-D12}]{\includegraphics[width=0.45\textwidth]{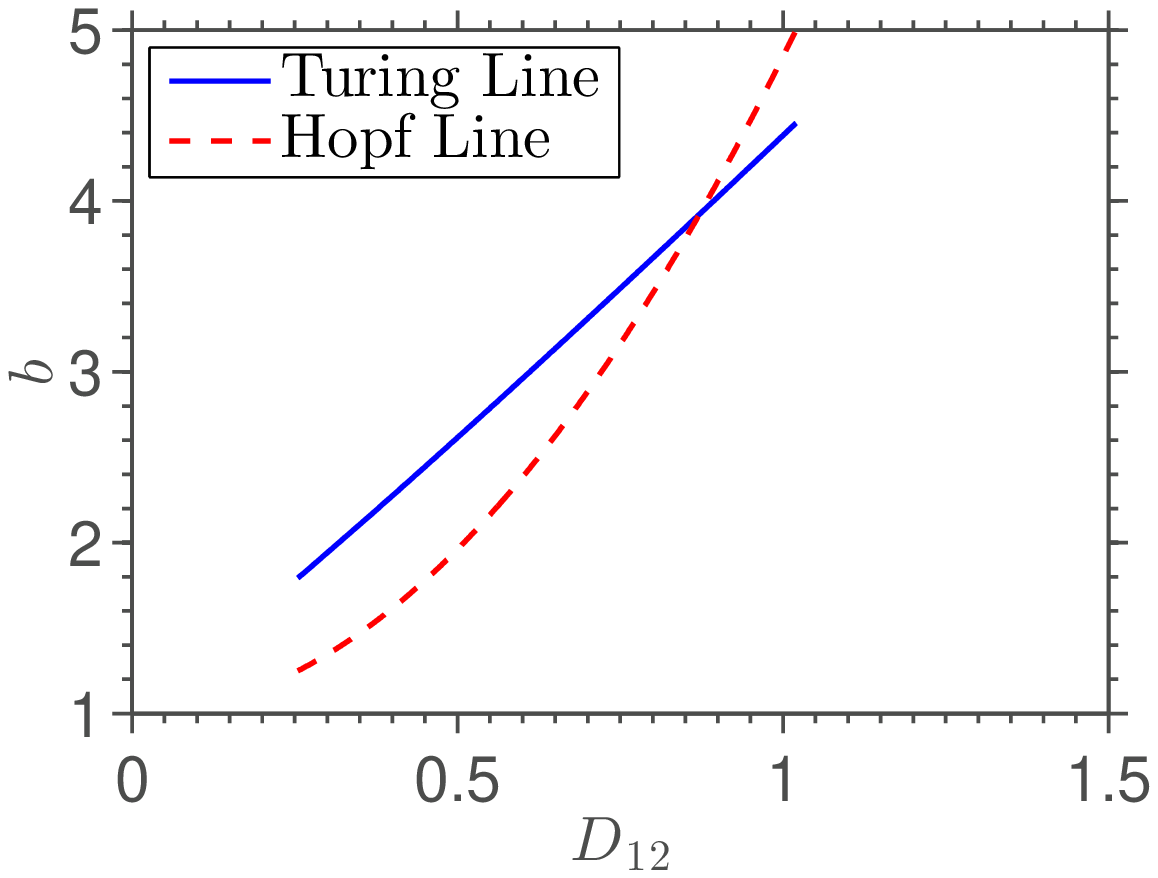}}\hfill
\subfigure[\label{b-D21}]{\includegraphics[width=0.45\textwidth]{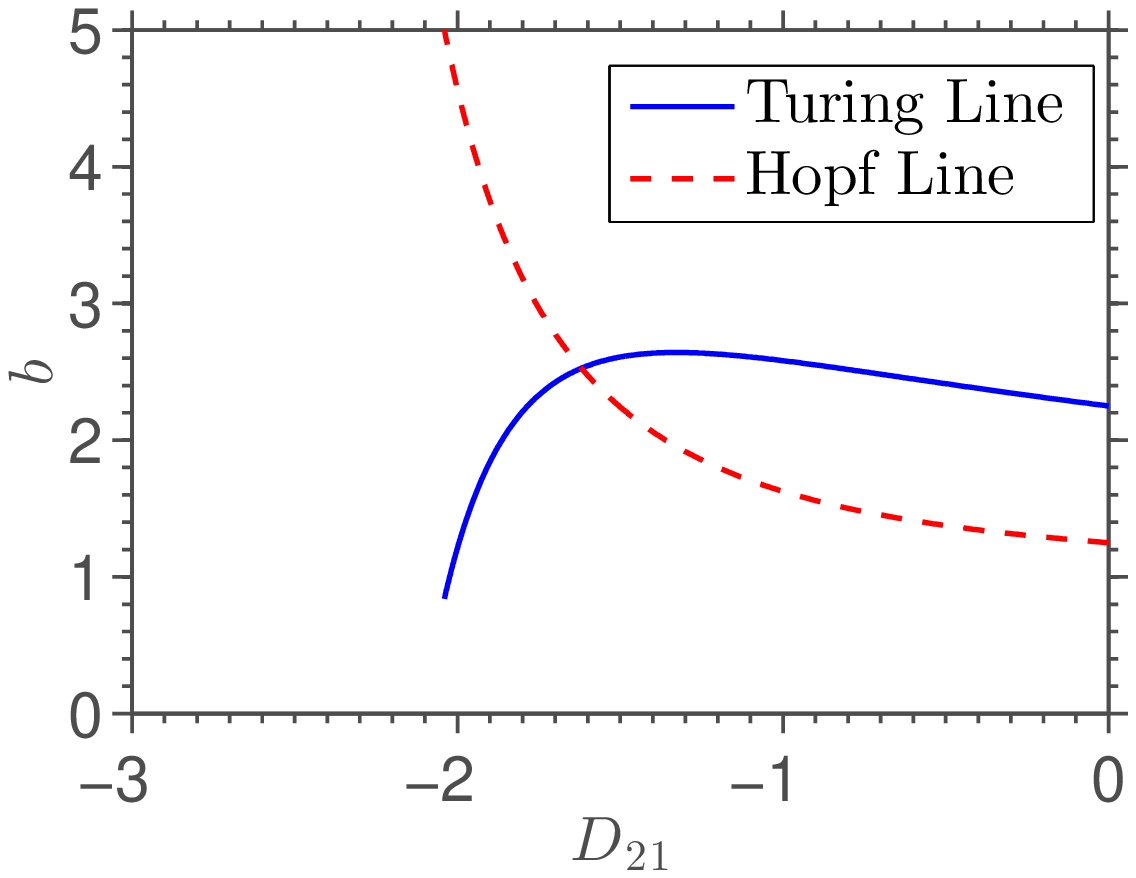}}
\caption{The solid line is the Turing line, and the dashed curve is the Hopf line in both FIG. \ref{parameter2_cd} and \ref{parameter_cd} in the presence of following diffusion coefficients: $D_{11}=D_{22}=1;D_{12}=0.51x_{0};D_{21}=-0.51y_{0}$. In FIG. \ref{parameter2_cd}, the point of intersection of Turing and Hopf line is shifted to lower values of a and b due to concentration-dependent cross diffusion coefficients. In FIG. \ref{parameter_cd}, the parameter, $D_{cr}=D_{12}[1+\frac{1}{a^2}]-D_{21}$ containing only cross diffusion coefficients show the effect of concentration dependent cross diffusion coefficients on Turing and Hopf line. FIG. \ref{b-D12} and \ref{b-D21} represent the individual effect of the cross diffusion coefficients $D_{12}$ and $D_{21}$, respectively. }
\label{fig:parameterspace_cd}
\end{figure*}

When cross diffusion has a linear dependence on the concentration, we can write the cross diffusion coefficients as $D_{12}=D_{12}x_{0}$ and $D_{21}=D_{21}y_{0}$\cite{Nkumar,Zemskov} with $x_{0}=\frac{k_1}{k_4}a$ and $y_{0}=\frac{k_2k_4b}{k_1k_3a}$ being the steady state concentrations of $X$ and $Y$ respectively for the spatially homogeneous system. As $A$ and $B$ are chemostatted species and kinetic rate constants are fixed at a particular value throughout the time of interest, $D_{21}$ and $D_{12}$ are effectively constant in this case also. From the FIG. \ref{parameter2_cd}, it is clear that the intersection of Turing and Hopf line is shifted to lower values of $a$ and $b$ due to this concentration dependence. In this context by taking only one non-zero cross diffusion coefficient at a time, we have shown the effect of individual cross diffusion coefficient on the Turing-Hopf line intersection in FIG. \ref{b-D12} and FIG. \ref{b-D21} for equal self diffusion coefficients. From FIG. \ref{parameter2_cd} and \ref{b-D21}, it is evident that modification of the Turing line due to concentration-dependent $D_{21}$ is comparable to the effect of concentration $a$ on the Turing line. From the FIG. \ref{parameter_cd}, \ref{b-D12} and \ref{b-D21}, one can conclude that $D21$ has stronger effect on the Turing line in the case of the Brusselator model. The more general concentration dependence of cross diffusion terms is beyond the scope here. 

We now consider three different values of $a$ for three different scenarios, i.e., $a=2.1$(Turing instability precedes Hopf instability), $a \approx 1.9438$ (COD2) and $a=1.8$( Hopf instability arises first) in subsequent studies. In FIG. \ref{parameter}, we explore Turing and Hopf line as a function of newly defined parameter, $D_{cr}=D_{12}[1+\frac{1}{a^2}]-D_{21}$ motivated by the fact that cross diffusion coefficients are present explicitly in this part and $D_{cr}$ appears in both eq. \eqref{kbamp} and  \eqref{kbphase} of amplitude and phase dynamics.

\begin{figure*}[tb!]
\centering     
\subfigure[\label{fig:ampt}]{\includegraphics[width=0.3\textwidth]{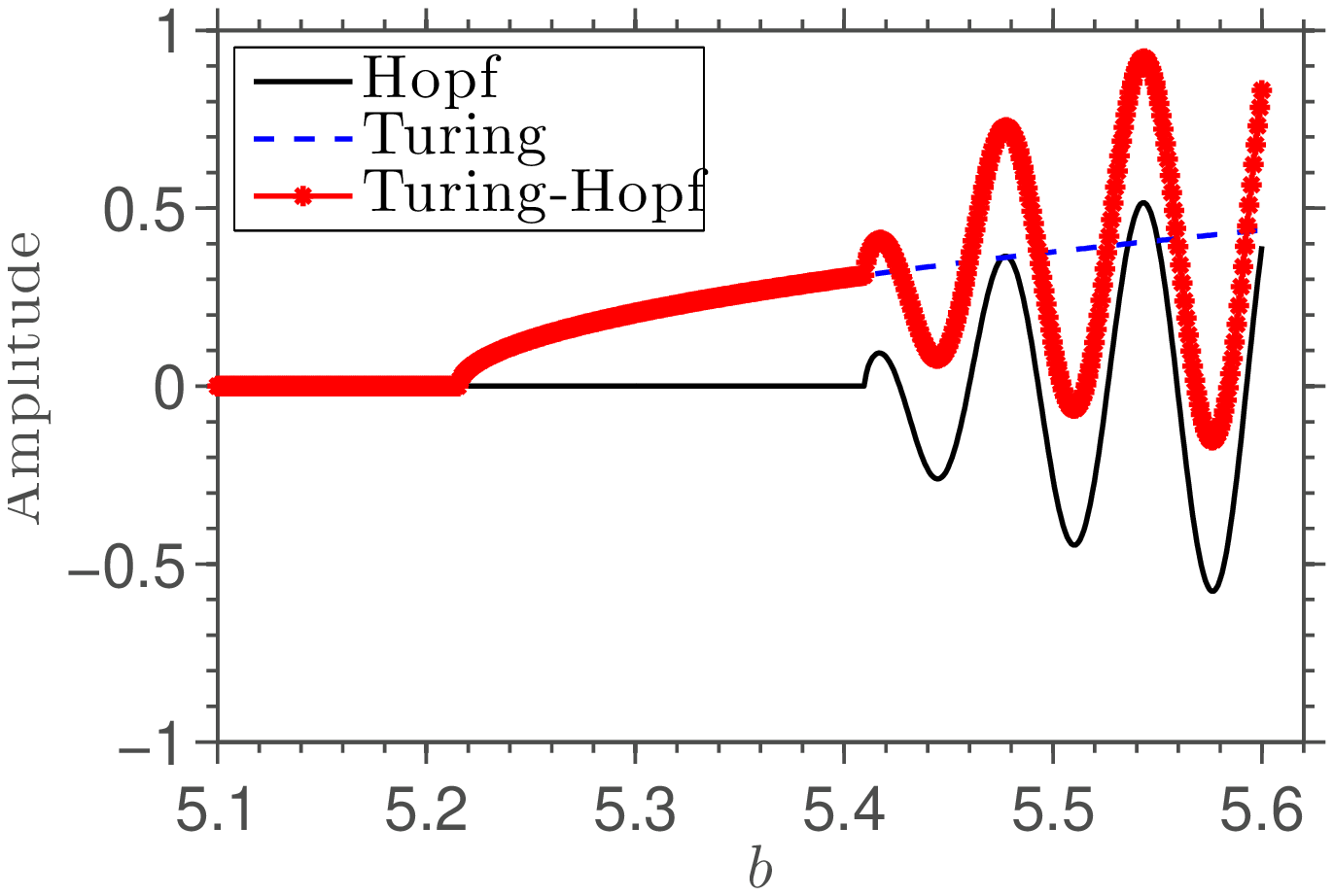}}\hfill
\subfigure[\label{fig:amp}]{\includegraphics[width=0.3\textwidth]{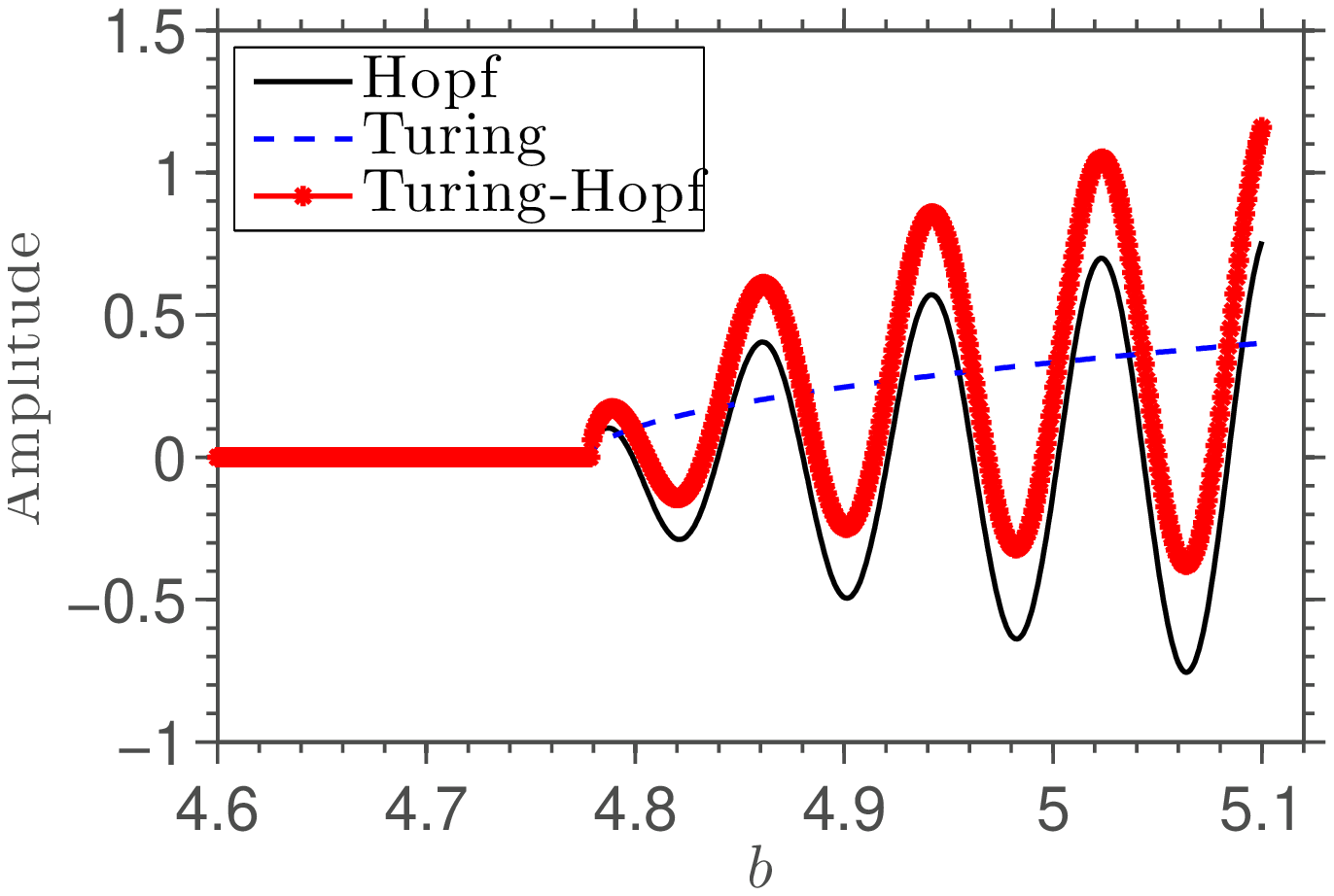}}\hfill
\subfigure[\label{fig:amph}]{\includegraphics[width=0.3\textwidth]{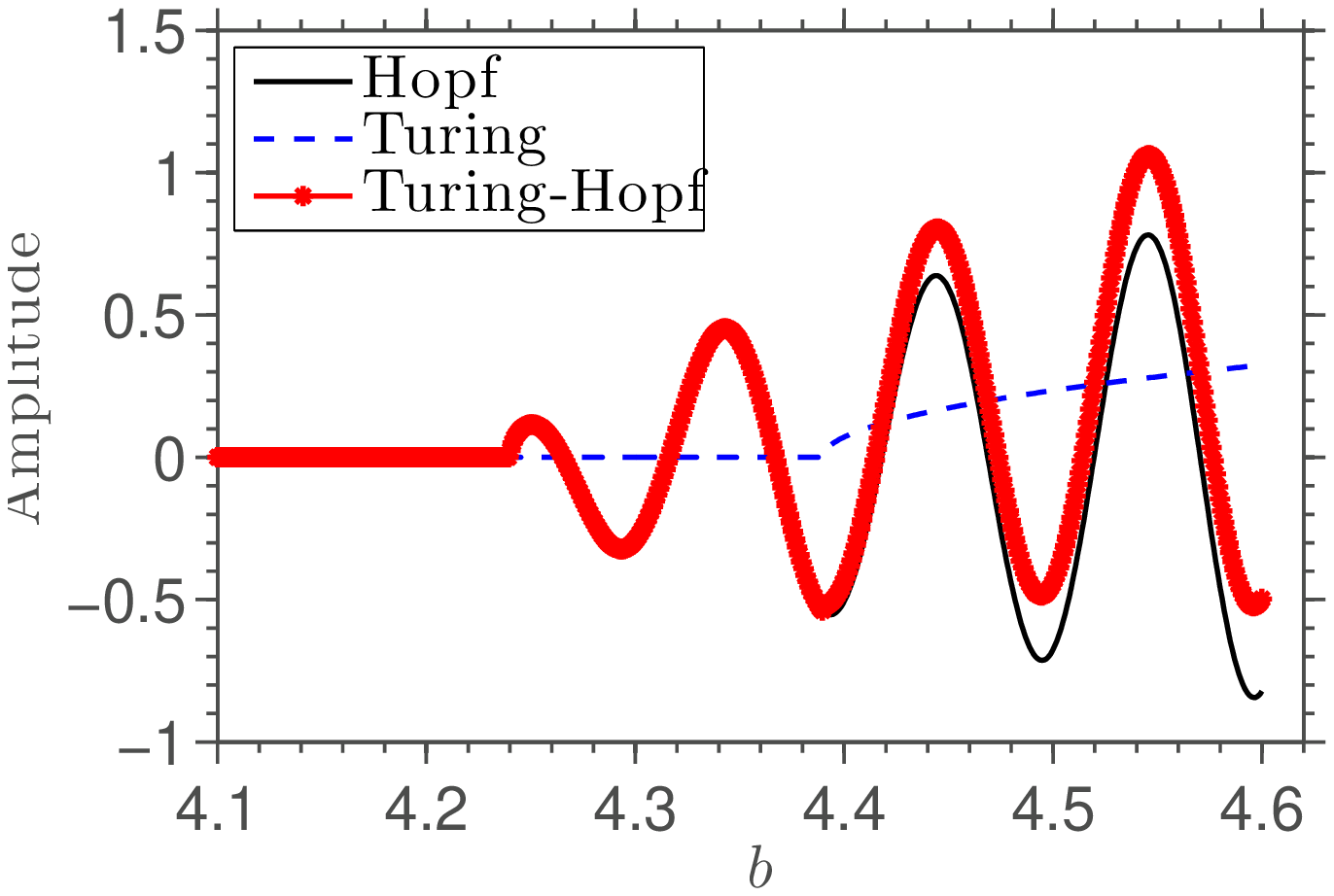}}
\caption{the solid black line is corresponding to Hopf amplitude derived analytically  as function of control parameter $b$ of the system at time $t=150$. The blue $'--'$ line refers to Turing amplitude obtained from the analytical amplitude equation of the Turing. While the red line plot with marker $'*'$ shows addition of Turing and Hopf amplitude. Three different scenarios  of Turing-Hopf interplay have been shown here: FIG. \ref{fig:ampt} for $a=2.1$, Turing first; FIG. \ref{fig:amp} for COD2-Turing and Hopf appear simultaneously at the same point in parameter space; FIG. \ref{fig:amph} for $a=1.8$, Hopf first. Here we have analyzed all three cases at a particular local point of the finite system of length $l=9.5$. This figures of amplitude will give lucid idea about local concentration profile in 1D Brusselator model in the parameter space of Turing-Hopf interplay and the effect of Hopf instability on the diffusion driven Turing instability.}
\label{amplitudeplot}
\end{figure*}

\begin{figure*}[htbp!]
\centering 
\subfigure[\label{fig.spatial}]{\includegraphics[width=0.3\textwidth]{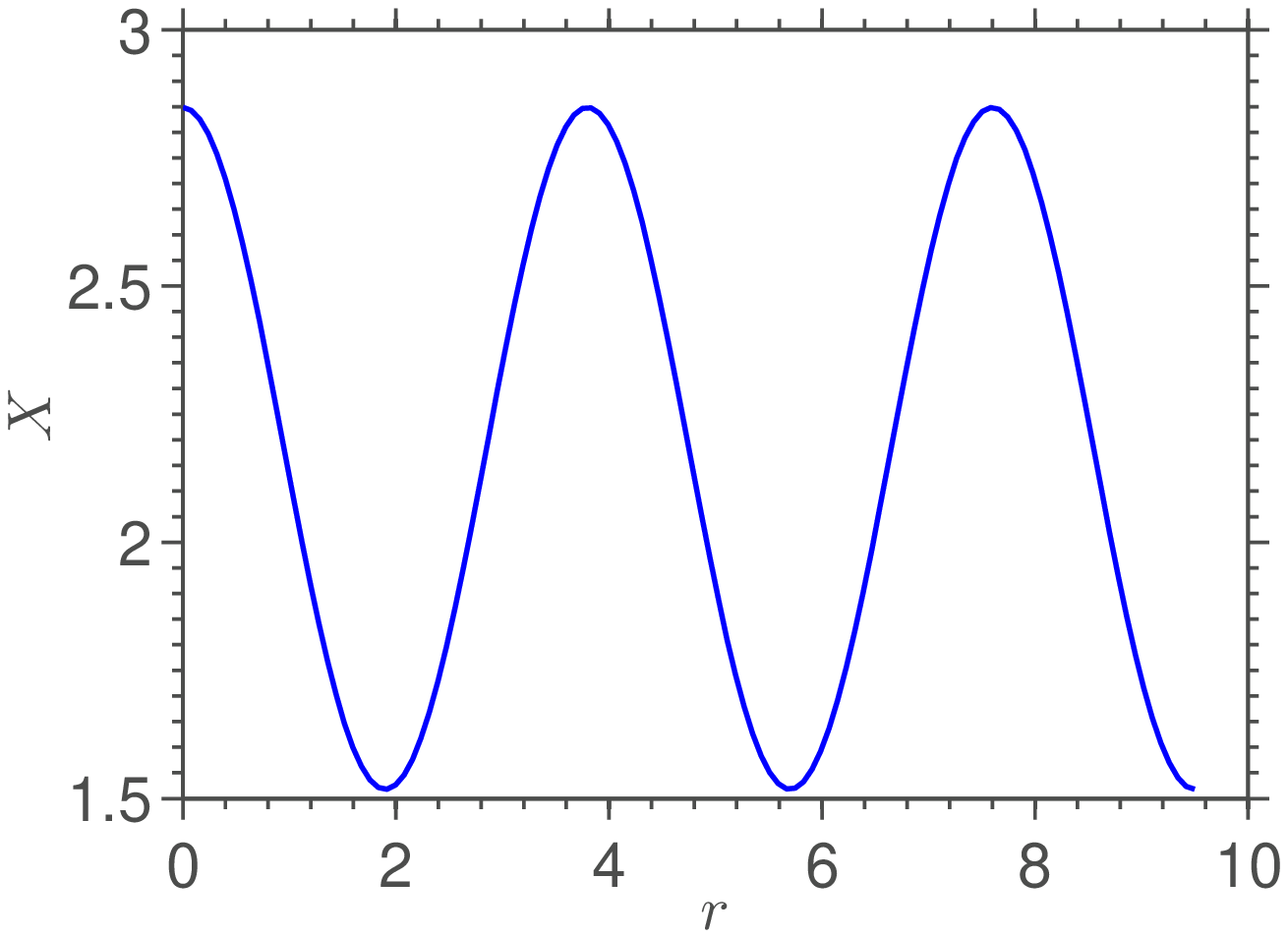}}\hfill
   \subfigure[\label{fig.temporal}]{\includegraphics[width=0.3\textwidth]{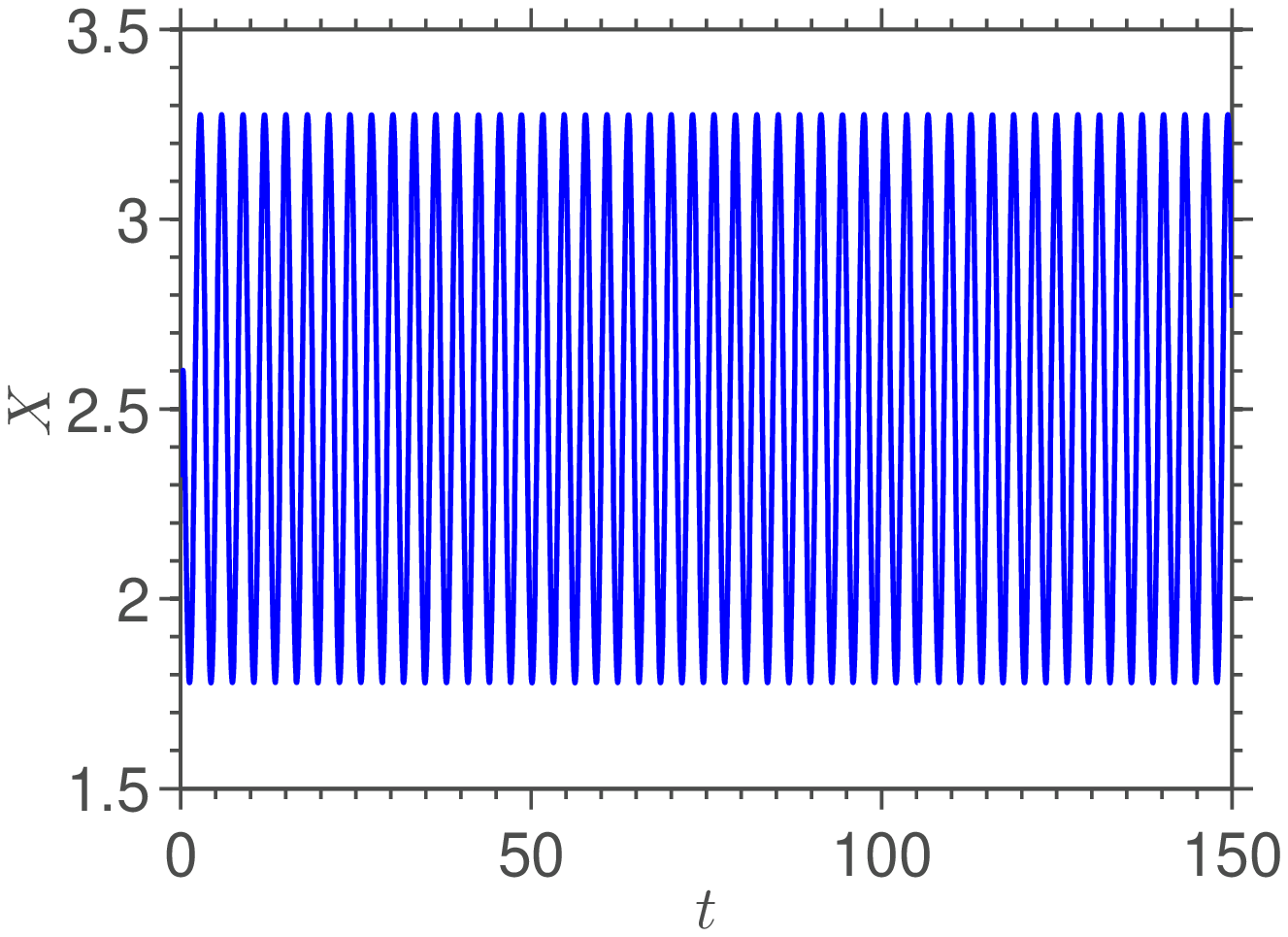}}\hfill
      \subfigure[\label{fig.powerspectrum}]{\includegraphics[width=0.3\textwidth]{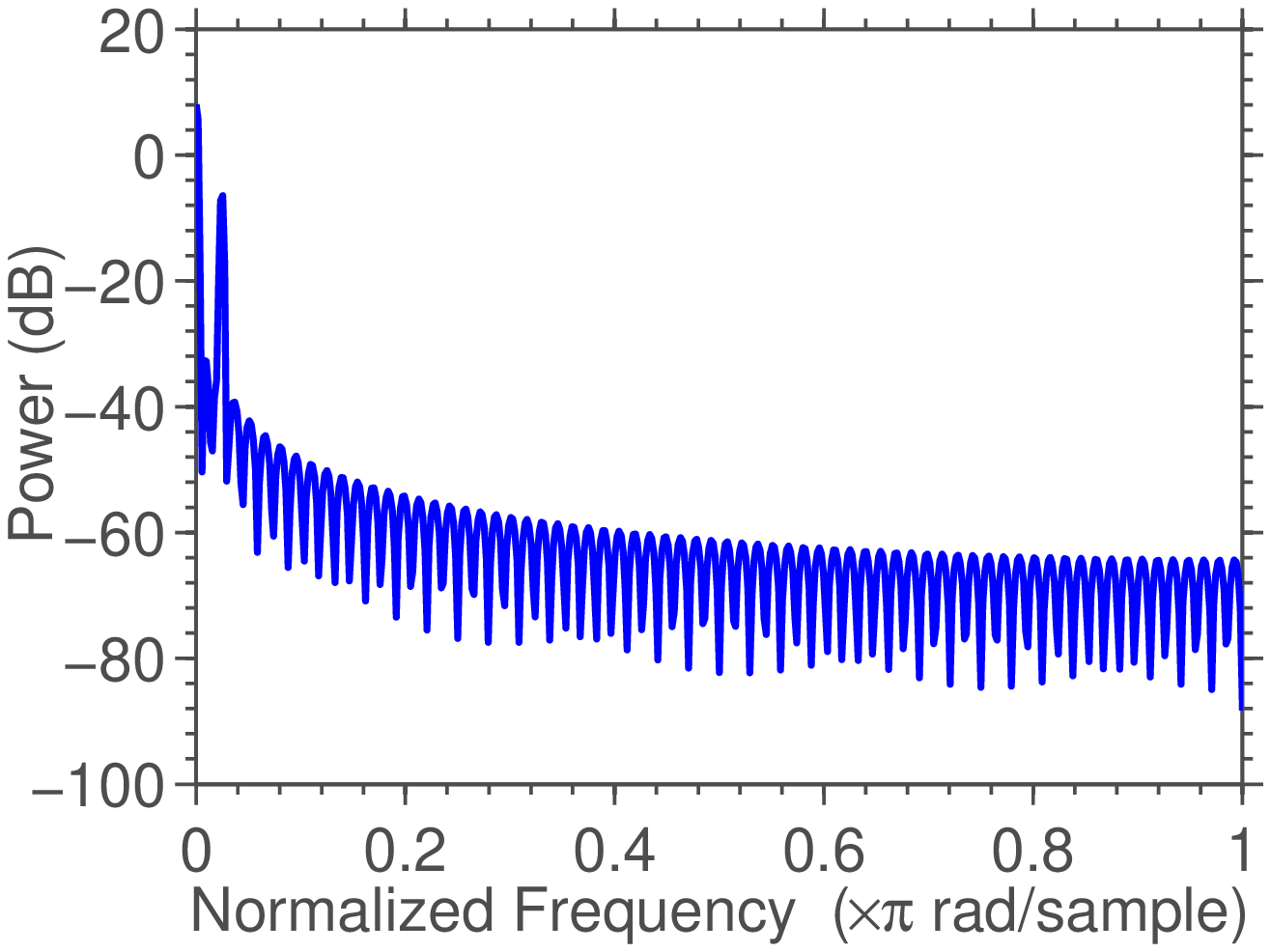}}
\caption{Spatial and temporal dynamics of concentration X  for the scenario when  Turing and Hopf mode arises simultaneously in a 1D Brusselator model with a system  size of $l=9.5$. Here $a=a_{TH}, b=4.9998$ and both self and cross diffusion is present. The spatial pattern in FIG.  \ref{fig.spatial} corresponds to a fixed time $t=150$. Whereas, temporal pattern in FIG. \ref{fig.temporal} corresponds to a particular point of the system. Existence of  these two structure owes to Turing instability and Hopf instability respectively. FIG. \ref{fig.powerspectrum}  shows power spectral density estimate of discrete-time concentration vector of species X(see FIG. \ref{fig.temporal}) obtained via Welch's method.}
\label{xrt}
\end{figure*}

In FIG. \ref{amplitudeplot}, we have shown the amplitude dynamics of Hopf and Turing with the aid of eq. \eqref{hopfamp} and \eqref{Turingamp}. As the control parameter, $b$ is changed through the critical values of Turing and Hopf instabilities, we can see how the oscillatory behavior of Hopf instability dominates over diffusion-driven Turing instability for steady state at a given point of the system at time $t=150$. Conversely, these figures also depict how the Turing instability modifies the oscillatory amplitude. As a consequence of this modification in oscillatory profile, the radius of corresponding limit cycle will also change. In another way, it shows effect of diffusion on the Hopf limit cycle indirectly through Turing instability. This amplitude profile renders clear idea about local concentration profile in the parameter space of the Turing-Hopf interplay. Above all, these figures are the measure of both Turing and Hopf instability at fundamental level.

In FIG. \ref{xrt}, the spatial and temporal profiles of the concentration for a given value of the control parameter show periodic behavior owing to Turing and Hopf instability, respectively. Spatial profile in FIG. \ref{fig.spatial} corresponds to a wave number close to the critical value of the Turing intrinsic critical wave number. Whereas, temporal oscillation in FIG. \ref{fig.temporal} has non zero normalized frequency as seen by the peak in power spectral density in FIG. \ref{fig.powerspectrum}. For the profiles in FIG. \ref{xrt}, we have considered only the scenario when Turing and Hopf instabilities arise simultaneously at a point in parameter space. For other two scenarios these profiles have  roughly the same features.

\begin{figure*}[tb!]
\centering 
 \subfigure[{\label{fig:EPRt}}]{\includegraphics[width=0.45\textwidth]{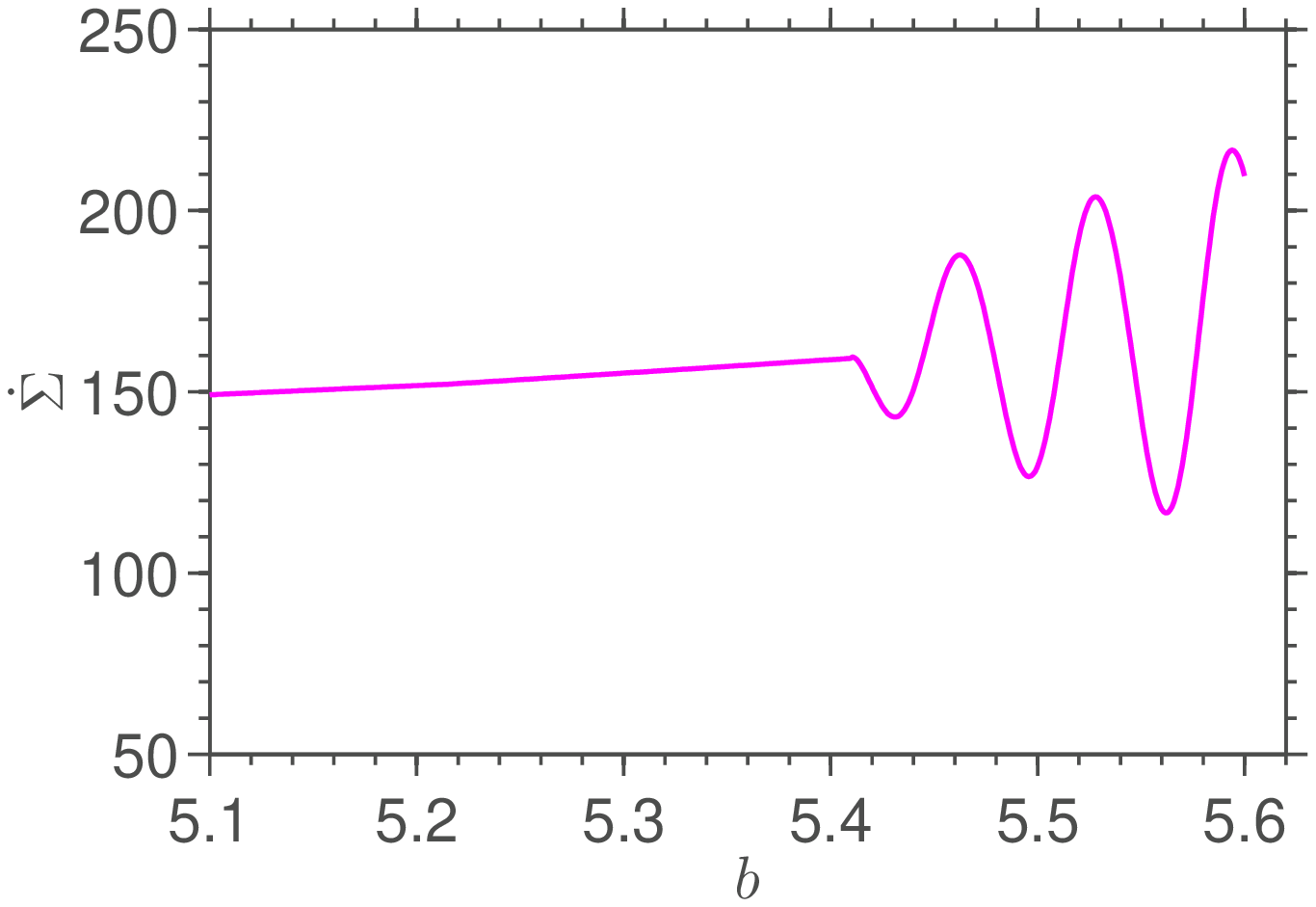}}\hfill
\subfigure[{\label{fig:xyft}}]{\includegraphics[width=0.45\textwidth]{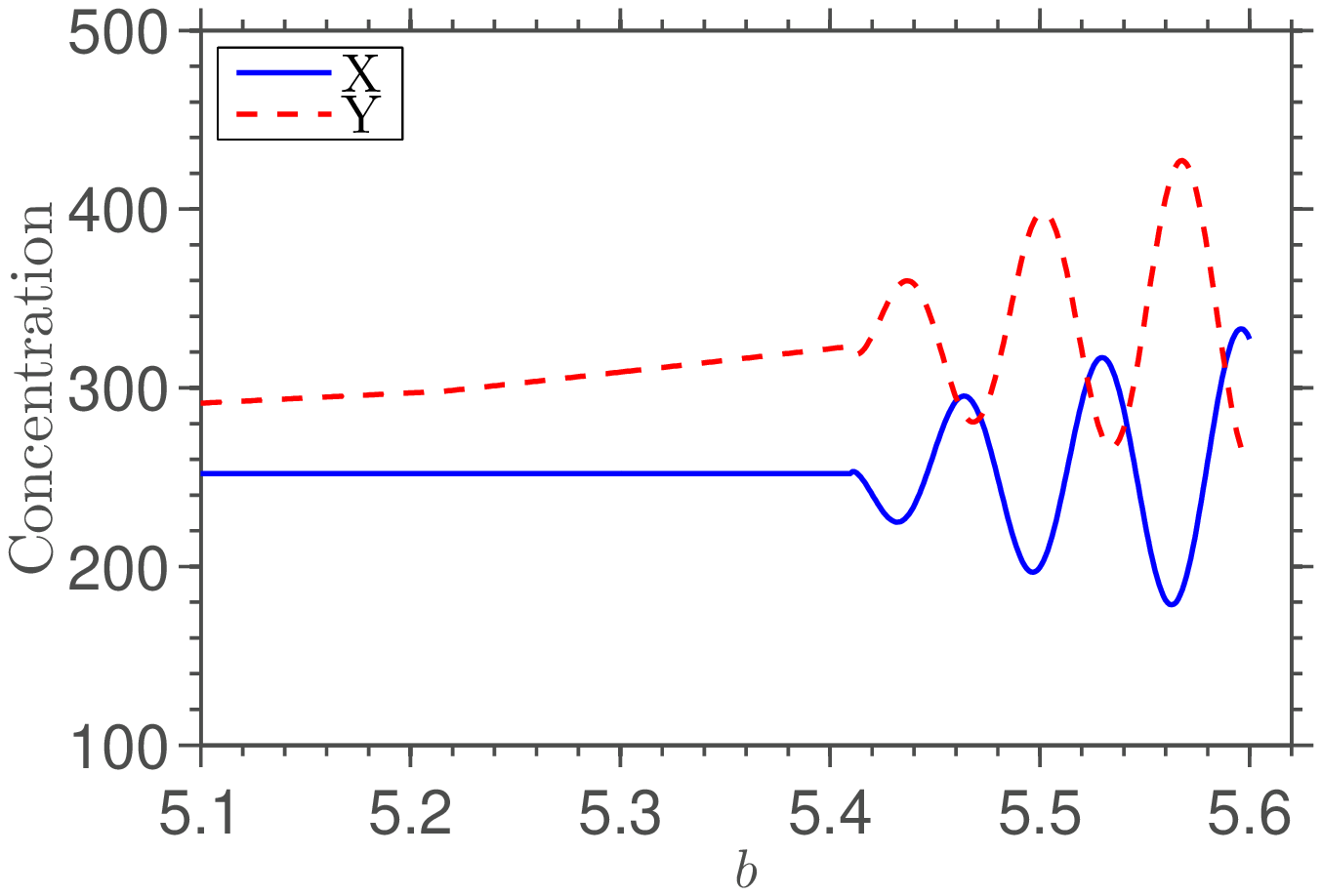}} 
\subfigure[{\label{fig:EPR}}]{\includegraphics[width=0.45\textwidth]{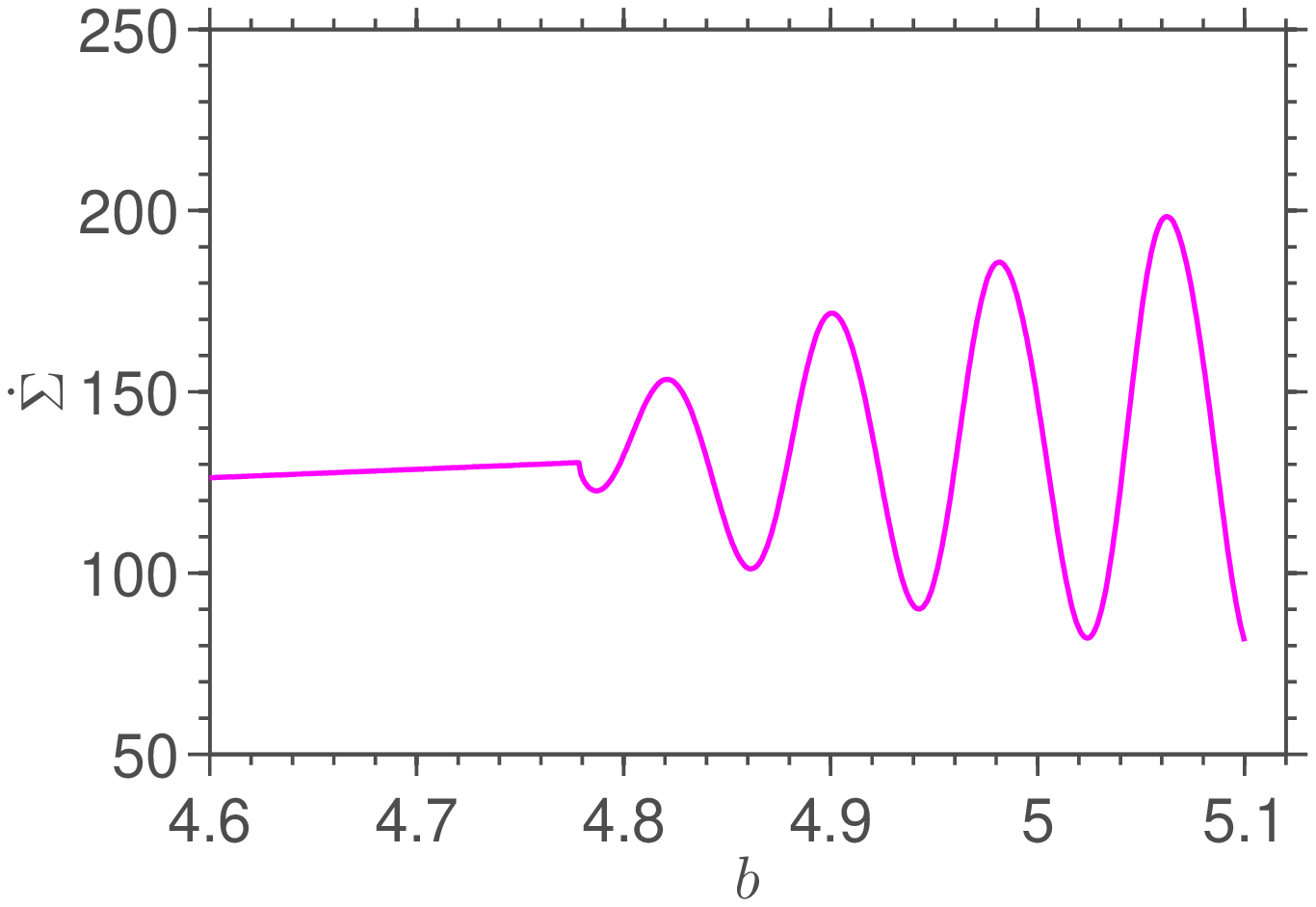}}\hfill
\subfigure[{\label{fig:xyf}}]{\includegraphics[width=0.45\textwidth]{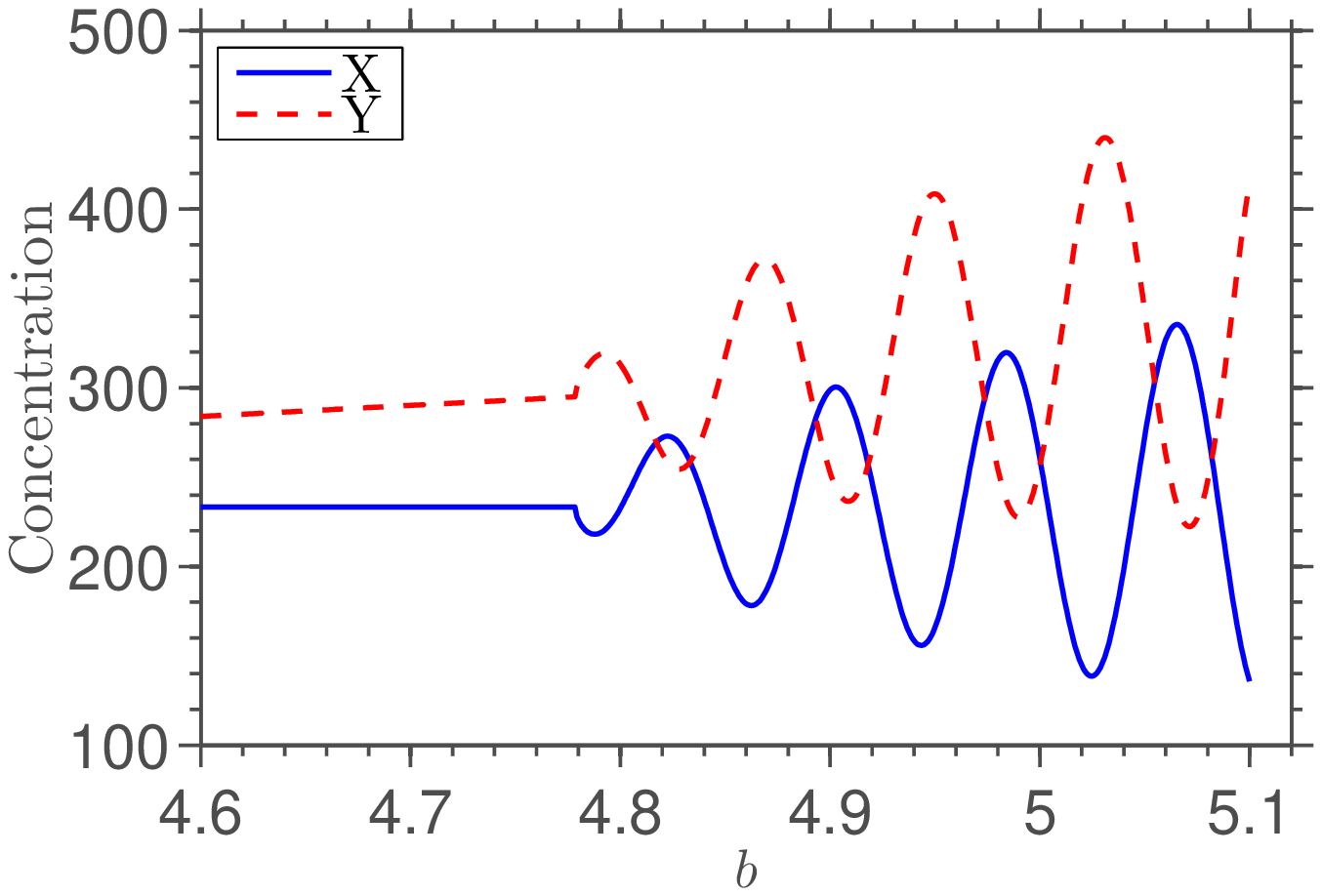}}
\subfigure[{\label{fig:EPRh}}]{\includegraphics[width=0.45\textwidth]{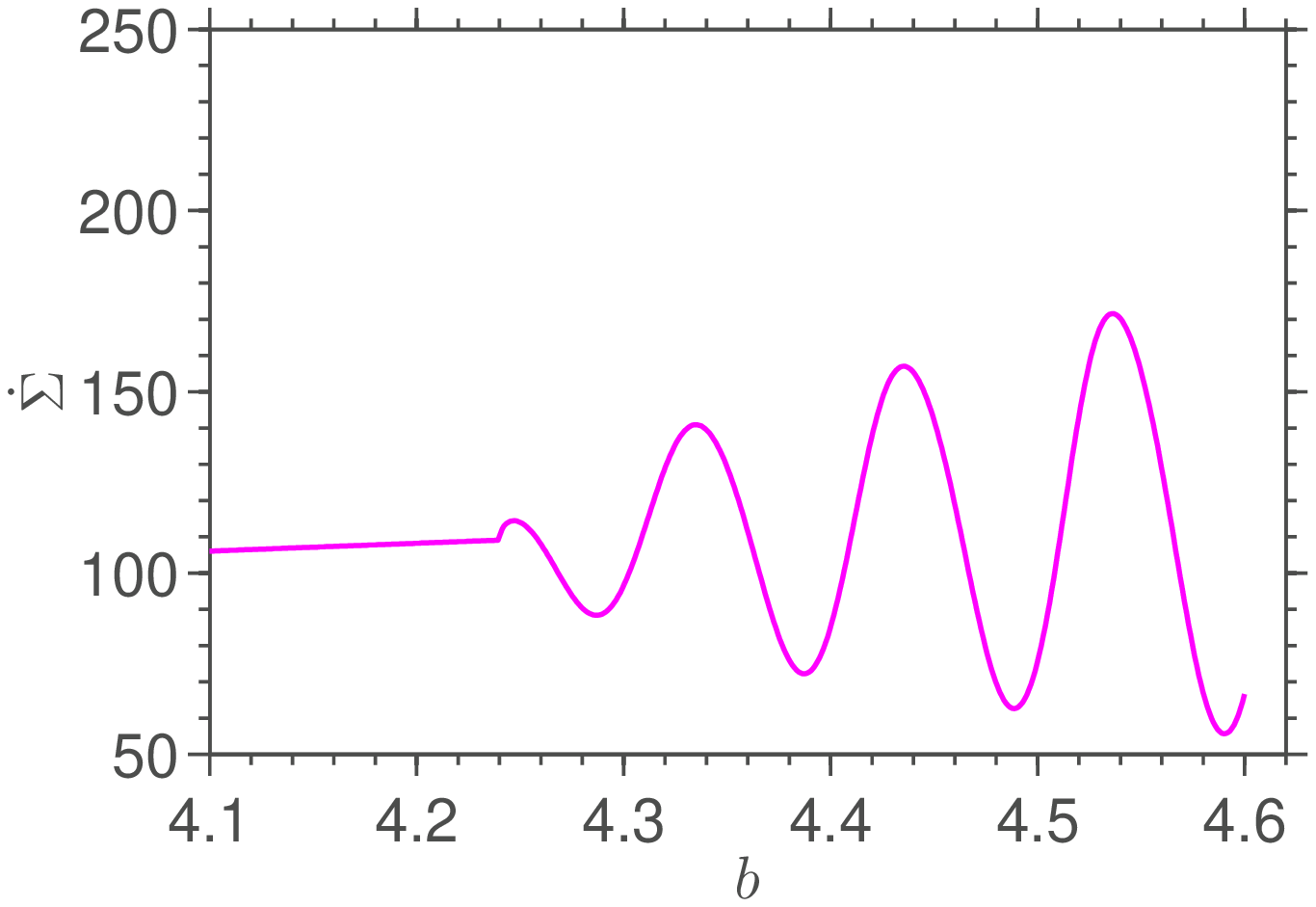}}\hfill
\subfigure[{\label{fig:xyfh}}]{\includegraphics[width=0.45\textwidth]{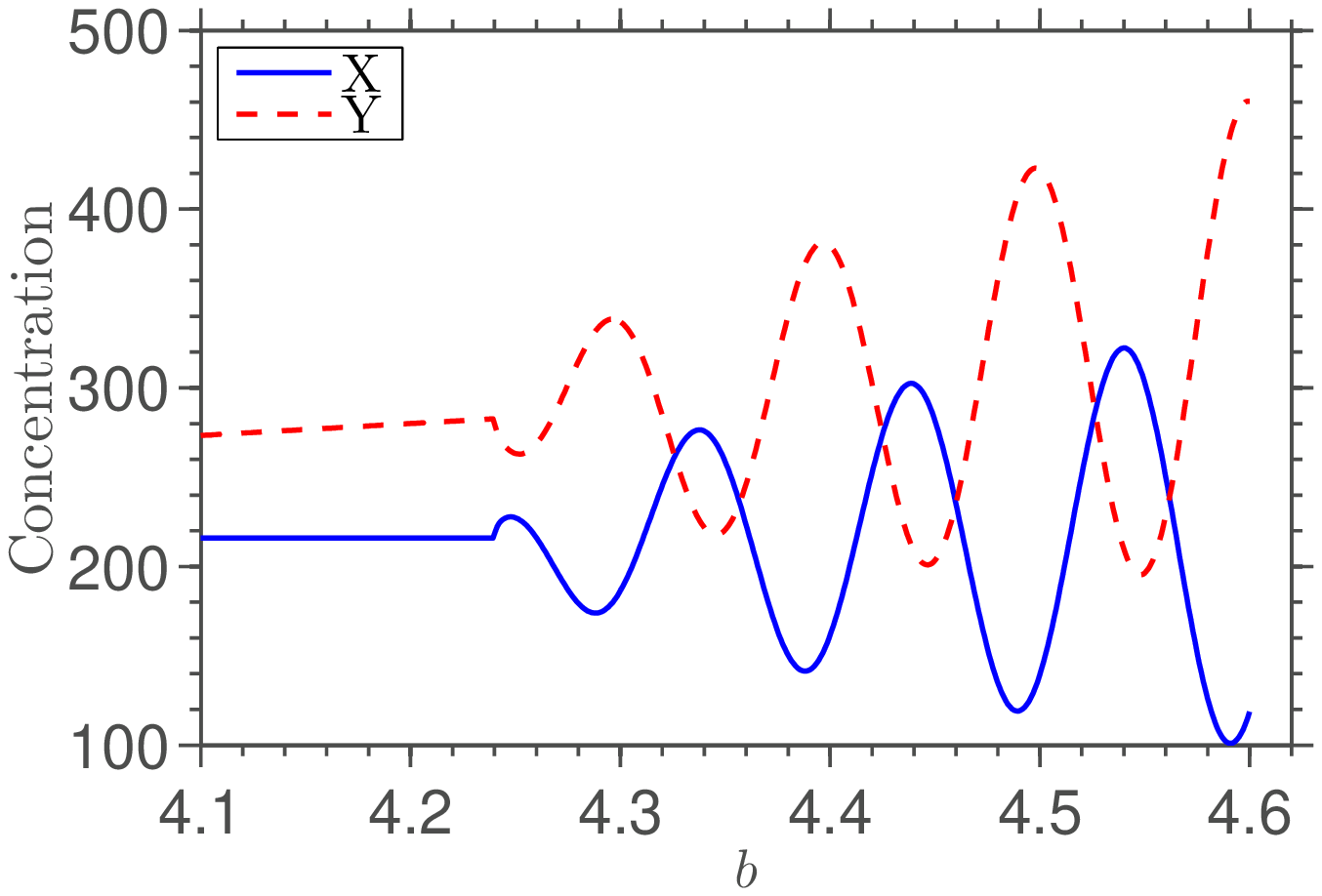}}
\caption{\label{fig:eprtxyf} Total entropy production(left column) as function of externally controlled parameter $b$ calculated analytically for a 1D Brusselator model of length $l=9.5$ at time, $t=150$ and absolute temperature, $T=300K$ for three different values of parameter $a$ leading to three different scenarios of Turing-Hopf interplay: FIG. \ref{fig:EPRt} and \ref{fig:xyft} for $a=2.1$, Turing first; FIG. \ref{fig:EPR} and \ref{fig:xyf} for COD2-Turing and Hopf appear simultaneously; FIG.  \ref{fig:EPRh} and \ref{fig:xyfh} for $a=1.8$,  Hopf first. Total entropy production expressed as the sum of entropy production rates due to diffusion and reaction parts. Global concentration field of intermediate species $X$ and $Y$ as a function of $b$ are shown on the right column. In all the three cases it is very apparent from the figures that entropy production rate is proportional to global concentration of $X$(or $Y$). For all the cases diffusion  coefficients are: $D_{11}=D_{22}=1;D_{12}=0.51;D_{21}=-0.51$ and reaction rate constants are $K_{-\rho}=10 ^{-4} << K_{\rho}=1.$(i.e., for weakly reversible case) }
\end{figure*}

\begin{figure*}[htbp!]
\centering 
\subfigure[{\label{fig:x3dft}}]{\includegraphics[width=0.3\textwidth]{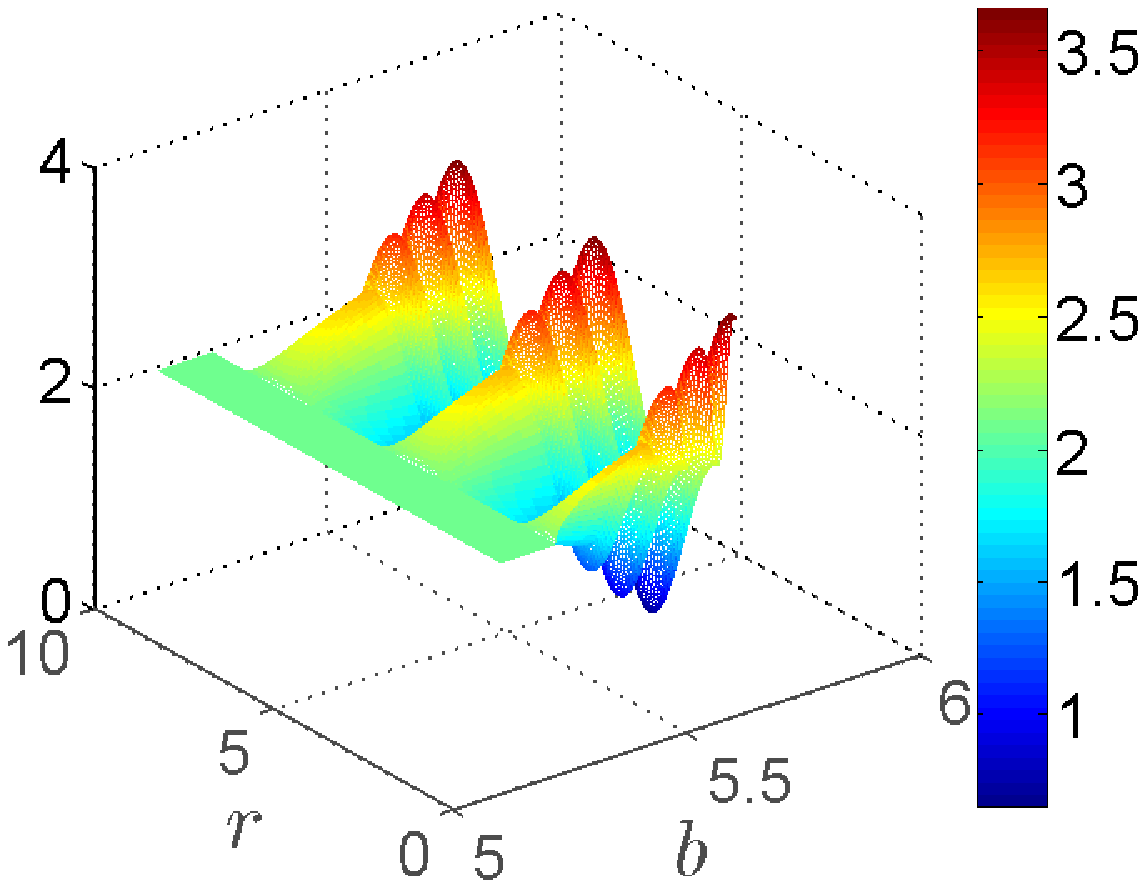}}\hfill
\subfigure[{\label{fig:x3df}}]{\includegraphics[width=0.3\textwidth]{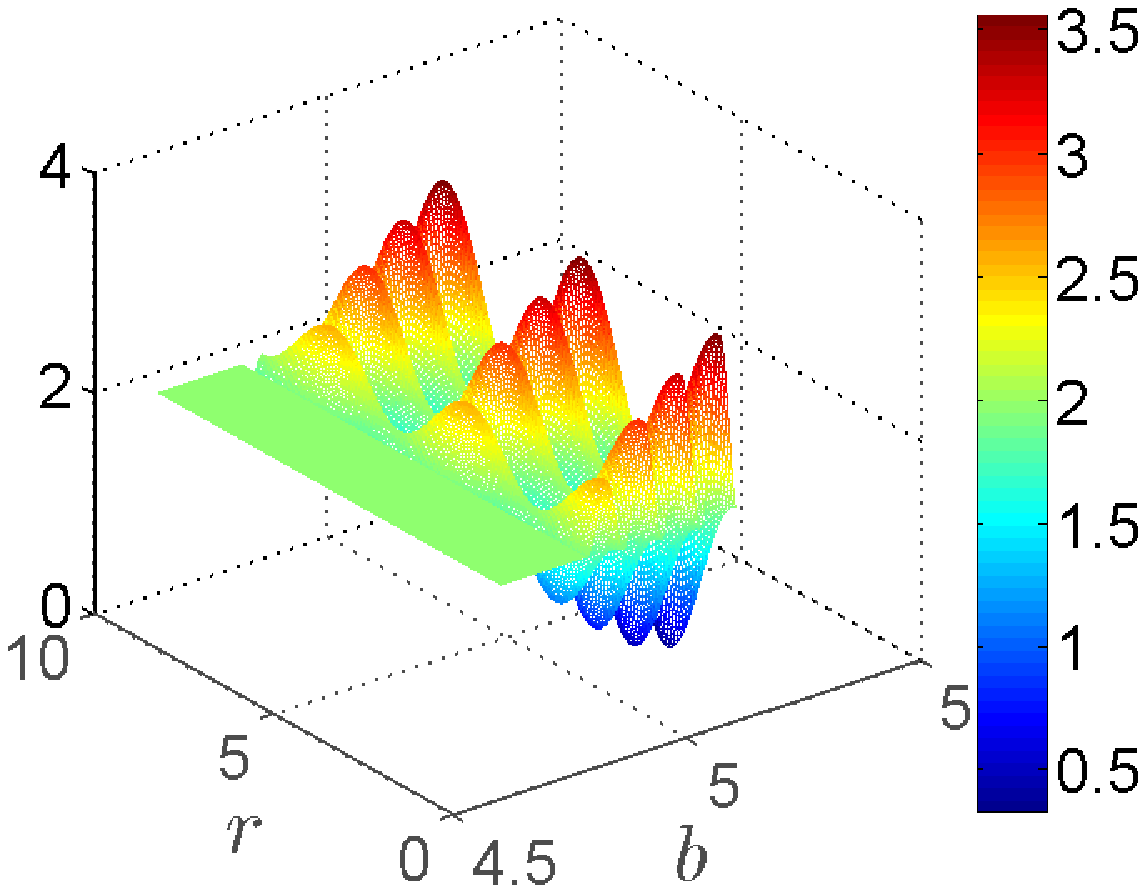}} \hfill
\subfigure[{\label{fig:x3dfh}}]{\includegraphics[width=0.3\textwidth]{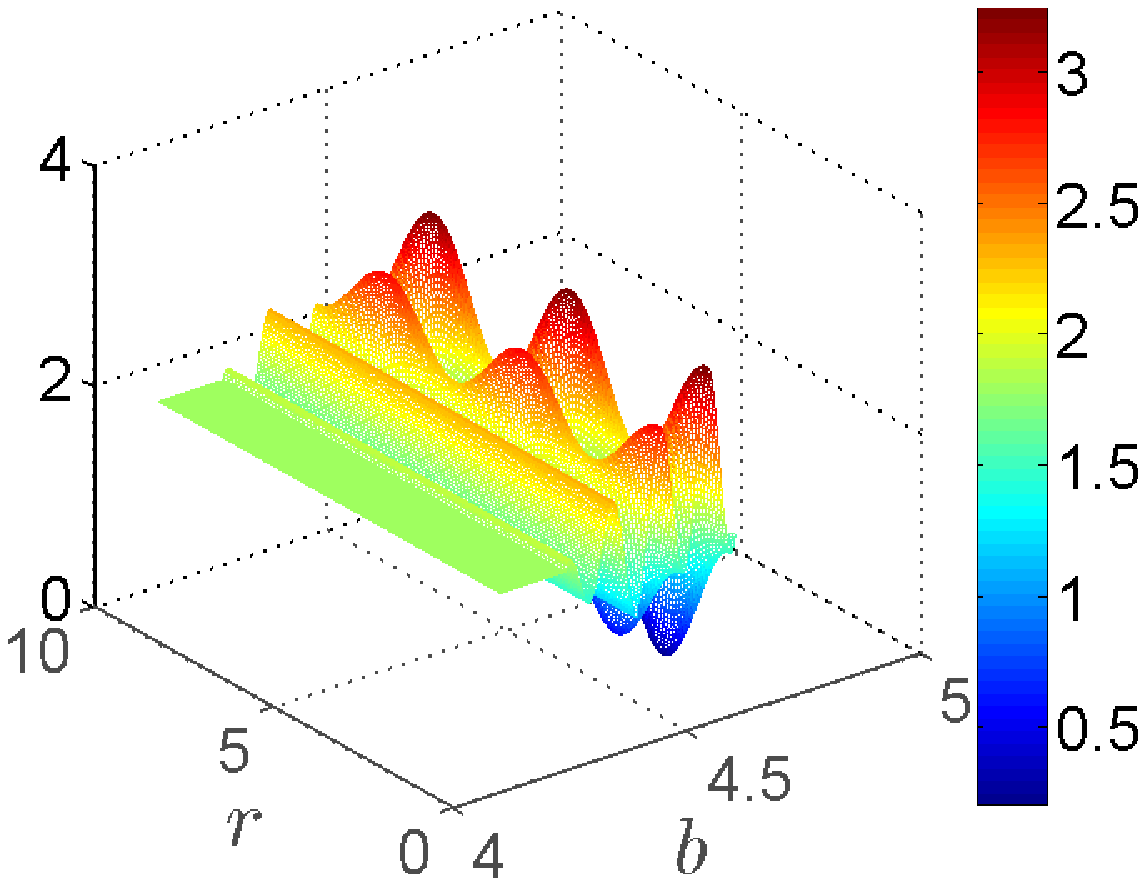}} \hfill
\subfigure[{\label{fig:xft}}]{\includegraphics[width=0.3\textwidth]{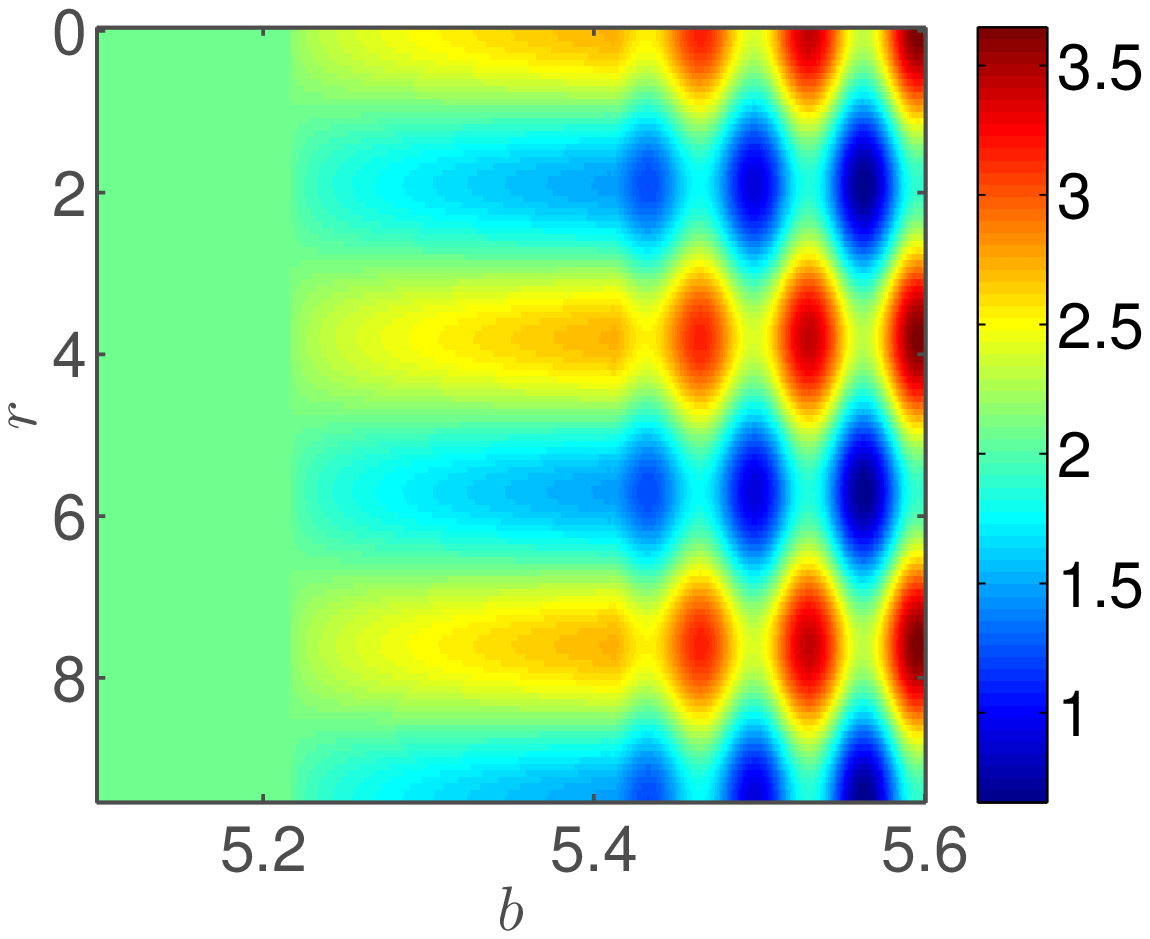}} \hfill
\subfigure[{\label{fig:xf}}]{\includegraphics[width=0.3\textwidth]{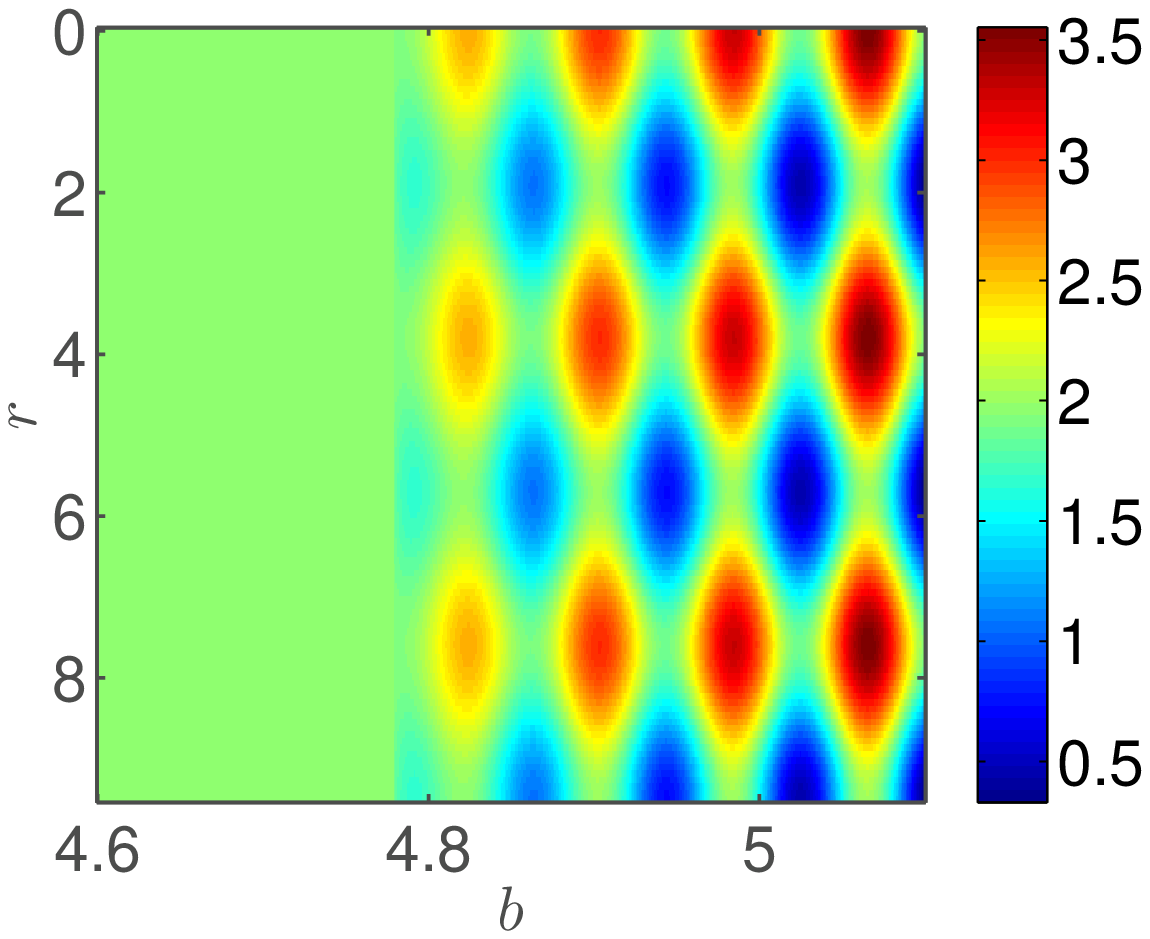}} \hfill
\subfigure[{\label{fig:xfh}}]{\includegraphics[width=0.3\textwidth]{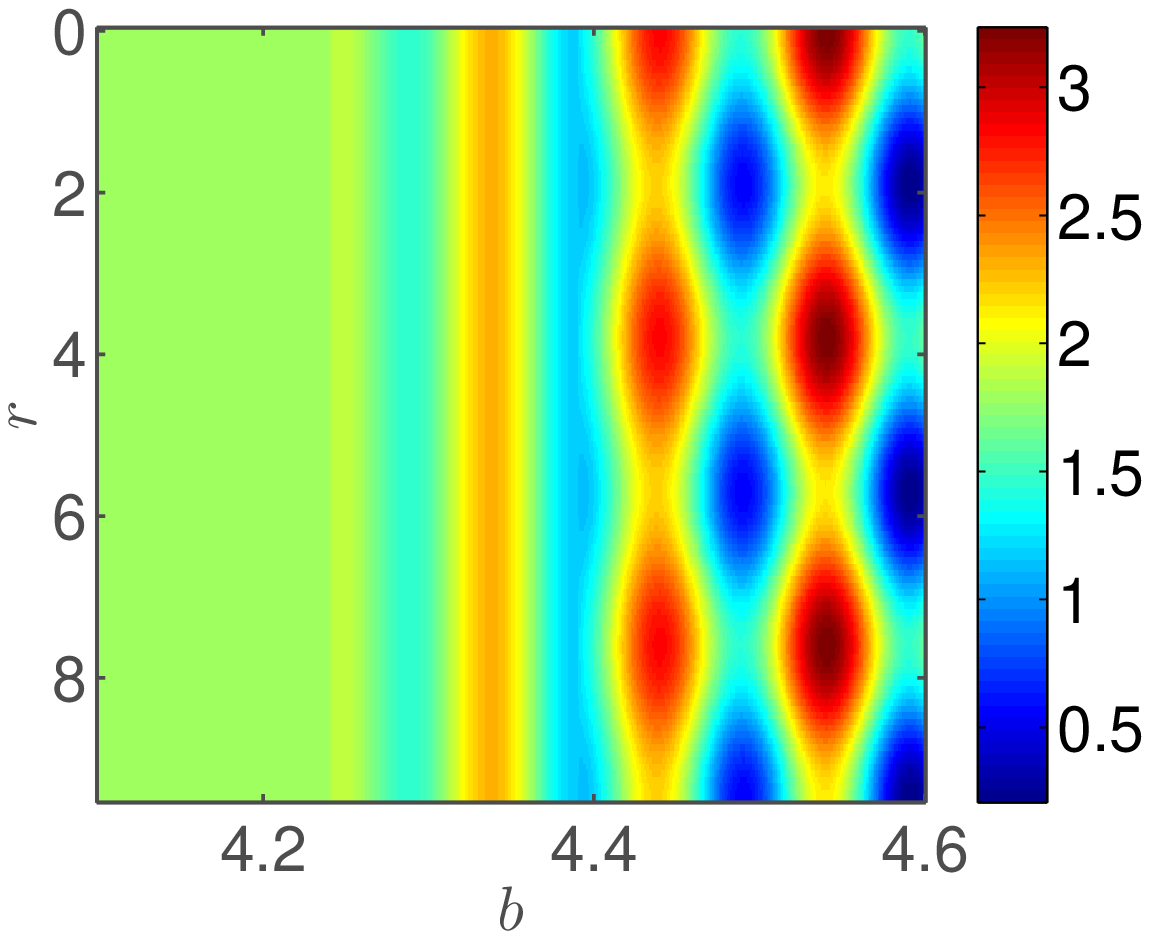}} 
\subfigure[\label{fig:EPBt}]{\includegraphics[width=0.3\textwidth]{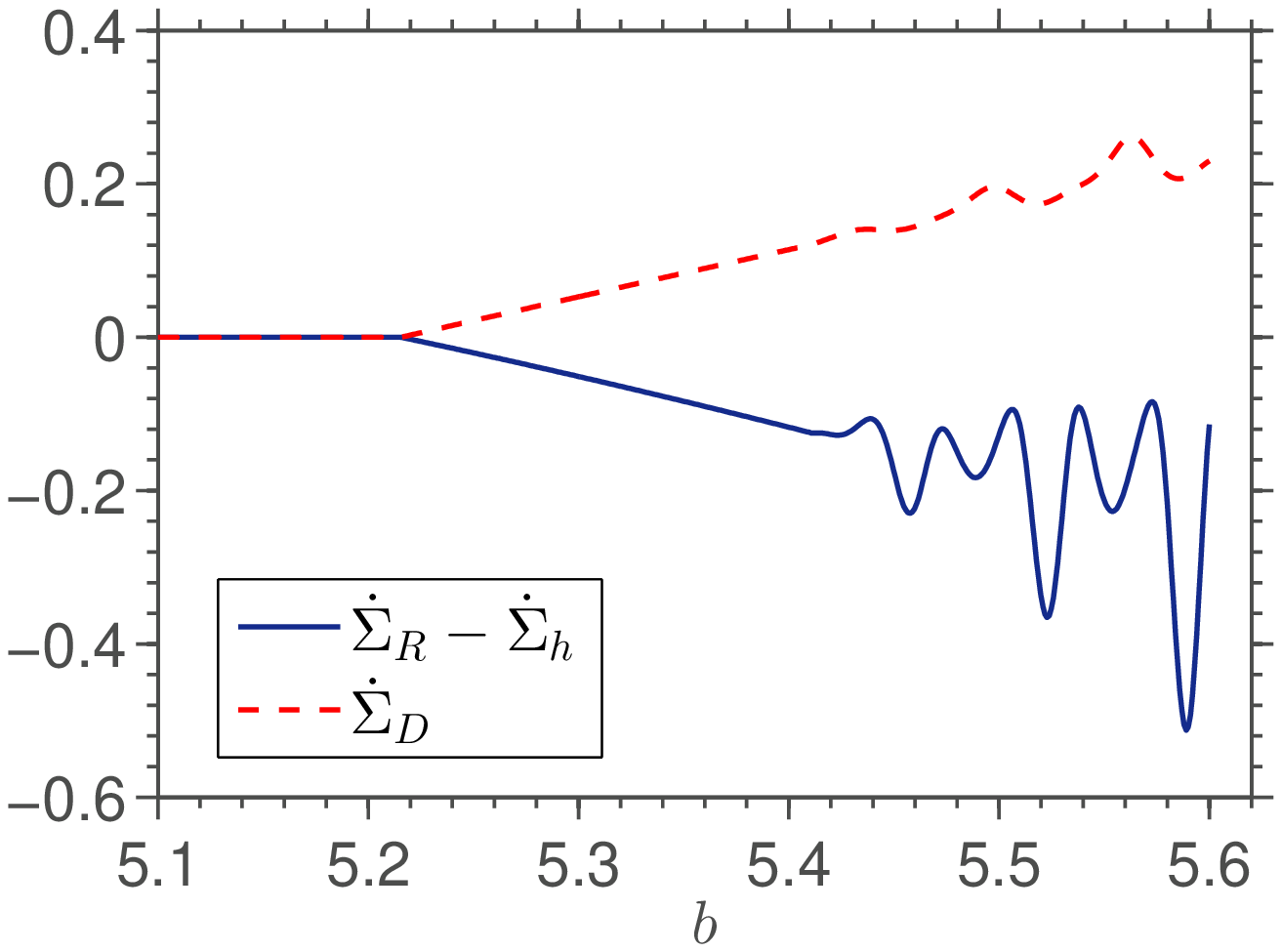} }\hfill
\subfigure[\label{fig:EPB}]{\includegraphics[width=0.3\textwidth]{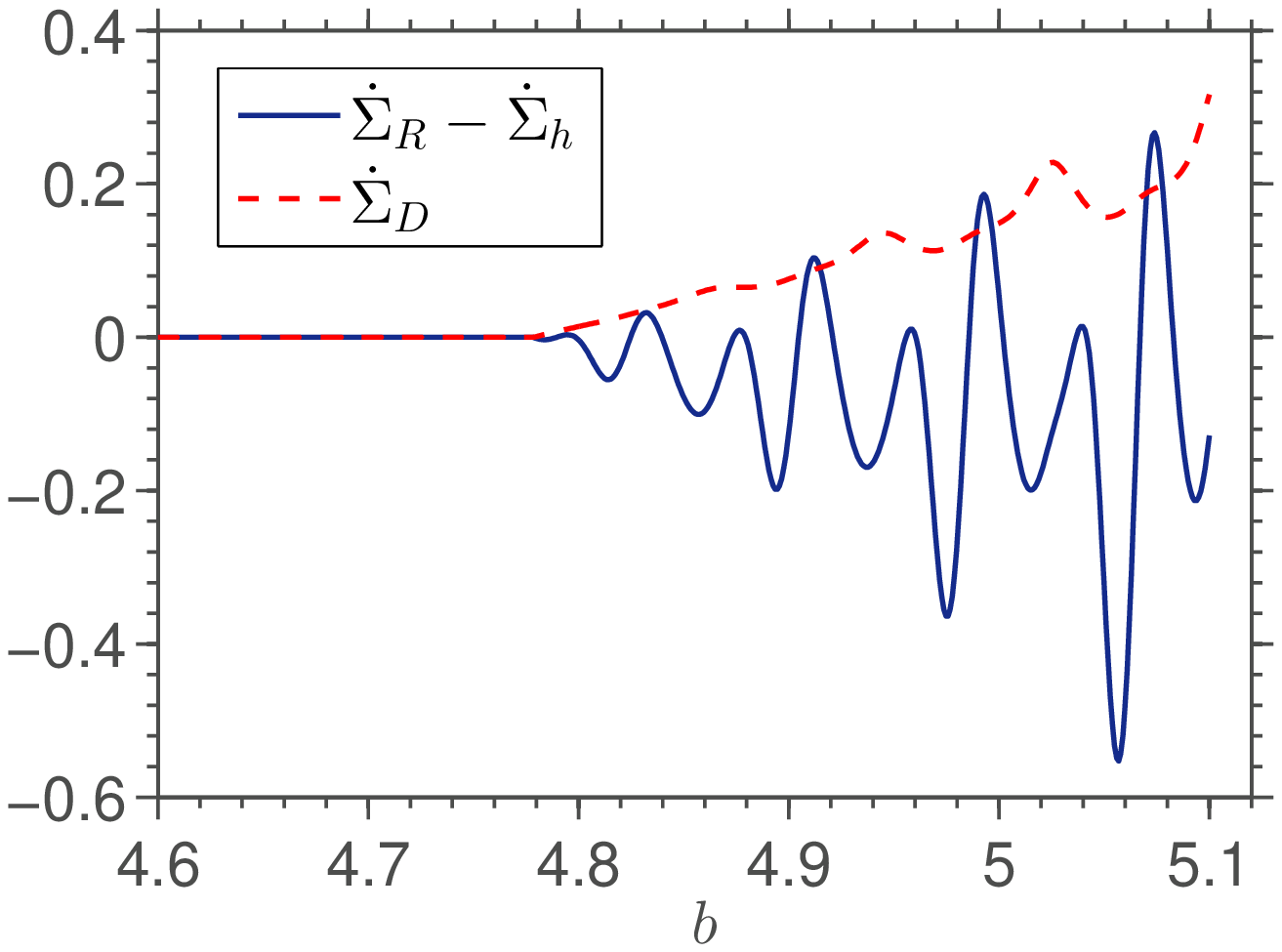}}\hfill
\subfigure[\label{fig:EPBh}]{\includegraphics[width=0.3\textwidth]{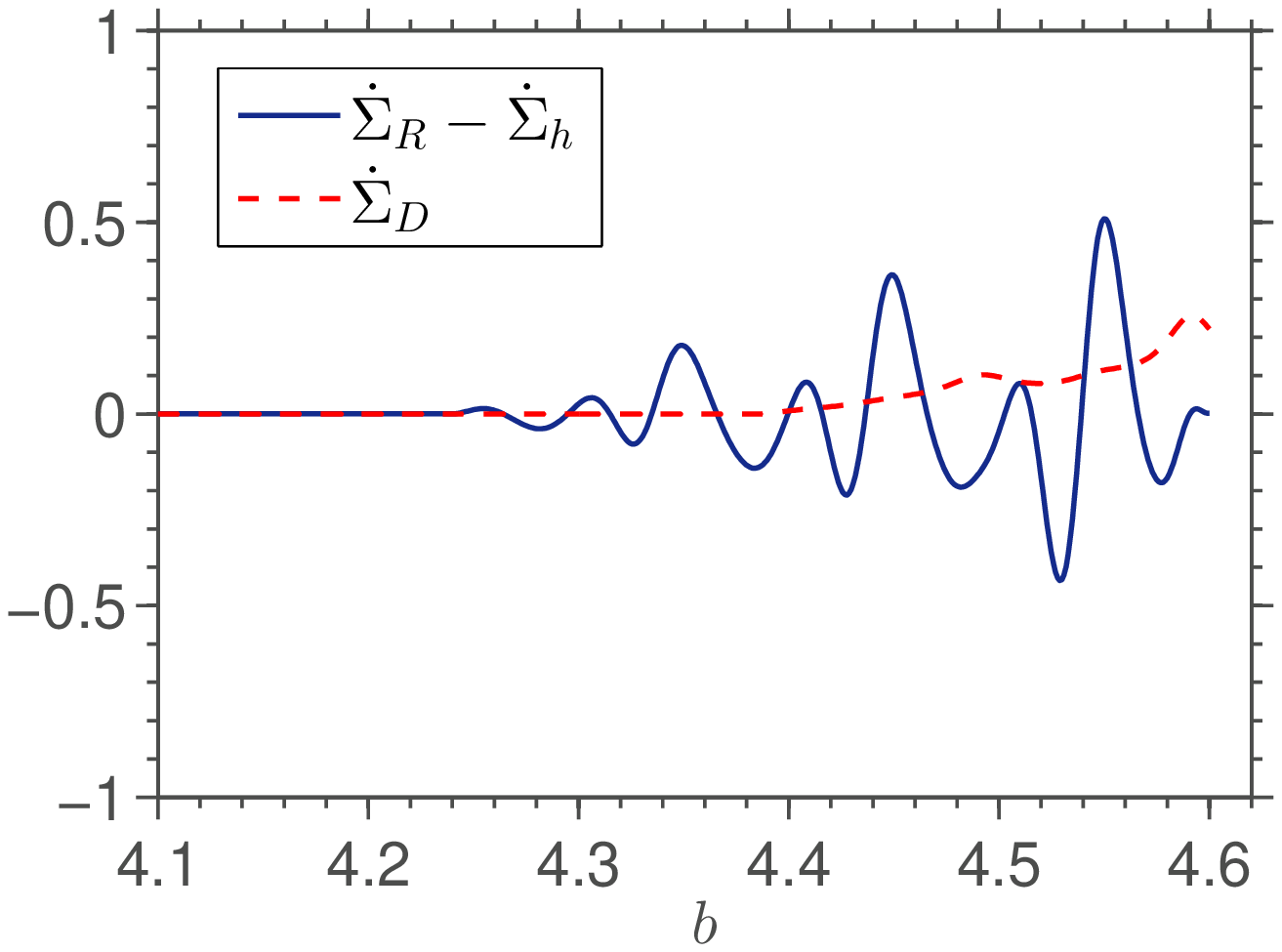}}
\caption{\label{compar}In the first row, 3d concentration fields of $X$ in the Brusselatator model of length $l=9.5$ as a function of externally controlled parameter, $b$ at $t=150$ and $T=300K$ for three different values of parameter $a$ leading to three different scenarios of Turing-Hopf interplay are shown(plots of $Y$ are similar). Here 'jet' colormap is used to show contrast in concentration field. Figures of the second row shows image of these concentration fields of $X$ . Extended spatial dimension is shown along the vertical axis. In the third row corresponding analytical result of EPR for reaction($\dot{\Sigma}_R$) and diffusion($\dot{\Sigma}_D$) as function of control parameter $b$ is presented. The solid blue lines correspond to the difference between reaction part and the homogeneous part of the reaction diffusion system and red $'--'$ lines refer to the diffusion part of entropy production rate. FIG. \ref{fig:x3dft}, \ref{fig:xft} and \ref{fig:EPBt} for $a=2.1$ Turing first; FIG. \ref{fig:x3df}, \ref{fig:xf} and \ref{fig:EPB} for COD2-Turing and Hopf appear simultaneously; FIG. \ref{fig:x3dfh}, \ref{fig:xfh} and \ref{fig:EPBh} for $a=1.8$  Hopf first.}
\end{figure*}

We have studied the response of the total entropy production rate due to the changes in reference chemostatted species, $b$ while another reference chemostatted species, $a$ remains constant. A nonzero total entropy production rate changes continuously and shows oscillatory response for all the three scenarios(as mentioned above) arise in Turing-Hopf interplay as shown in FIG. \ref{fig:EPRt},\ref{fig:EPR} and \ref{fig:EPRh}. Comparison between profiles of global concentration of activator(or inhibitor) on the right column of FIG. \ref{fig:eprtxyf} and corresponding total EPR on the left column of the same figure reveals that the total entropy production rate is quantitatively  proportional to the total concentration of activator(or inhibitor) in the  reaction diffusion system. Moreover, they are  showing qualitatively similar dynamics for all three cases. In other words, entropy production rate reflects the  global dynamics of  reaction diffusion system concentration arising from the Turing-Hopf interplay.  This result simply implies that the entropy production rate of a dissipative system can measure the pattern formation  quantitatively as well as qualitatively.

\begin{figure*}[tb!]
\centering 
\subfigure[{\label{fig:sggt}}]{\includegraphics[width=0.415\textwidth]{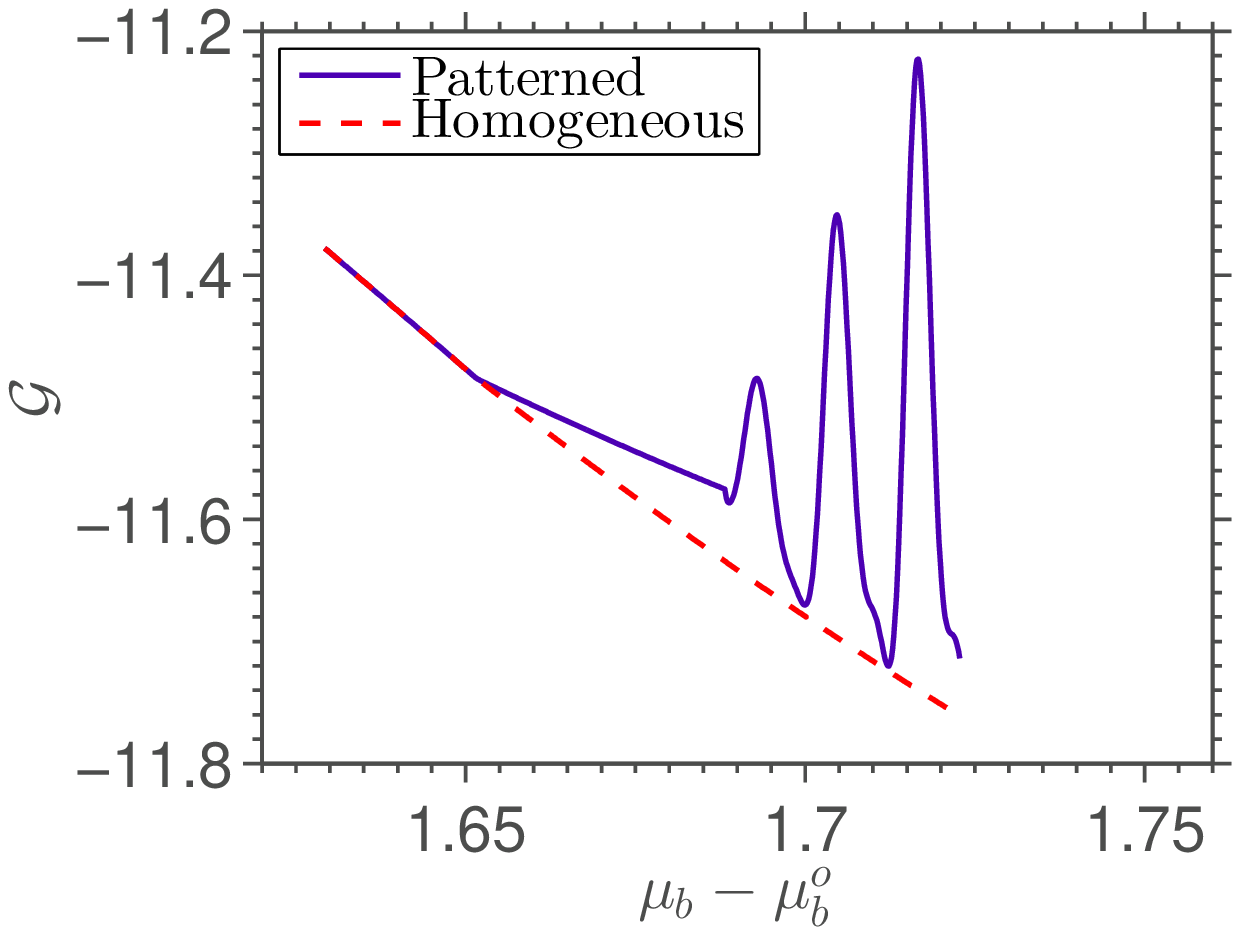}}\hfill
\subfigure[{\label{fig:slopet}}]{\includegraphics[width=0.415\textwidth]{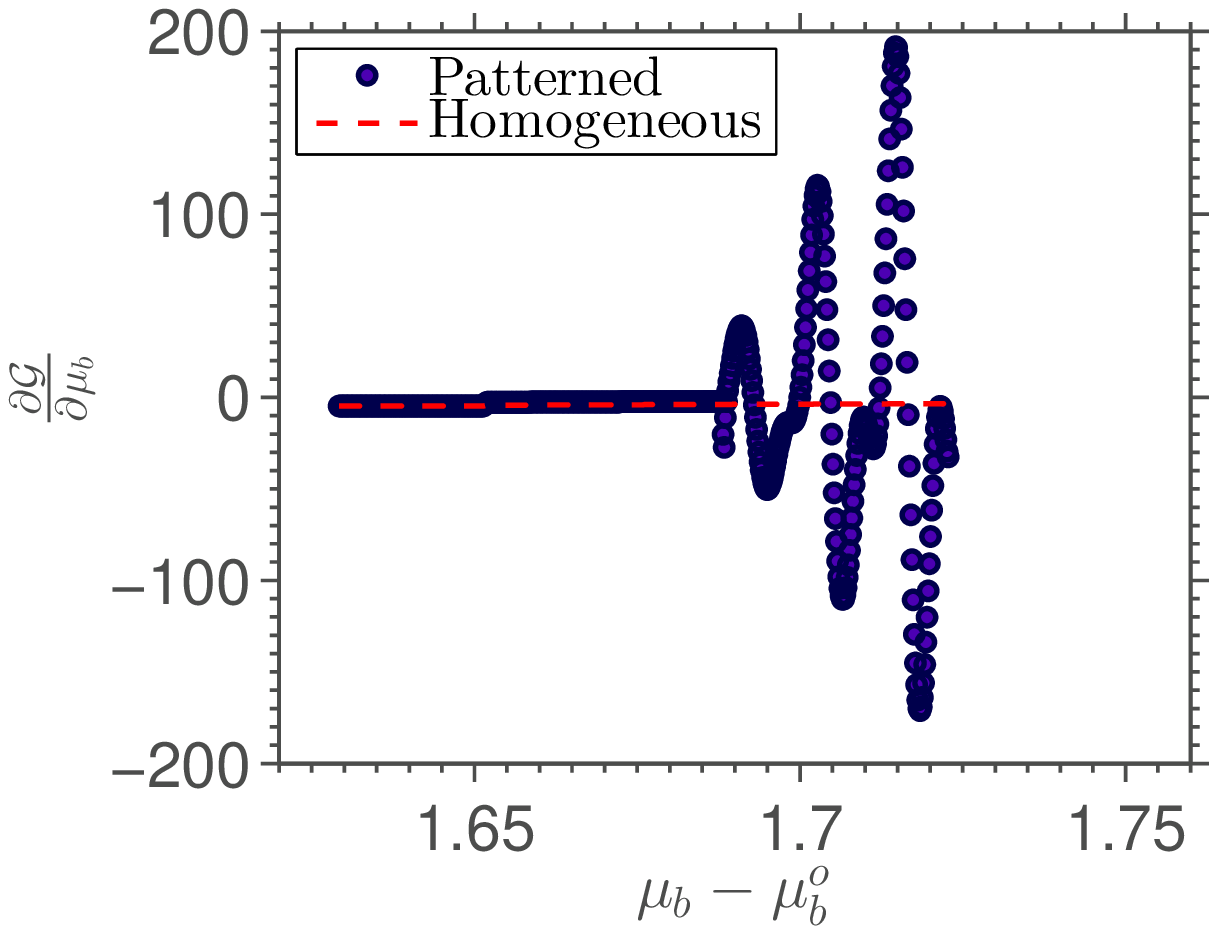}} 
\subfigure[{\label{fig:sgg}}]{\includegraphics[width=0.415\textwidth]{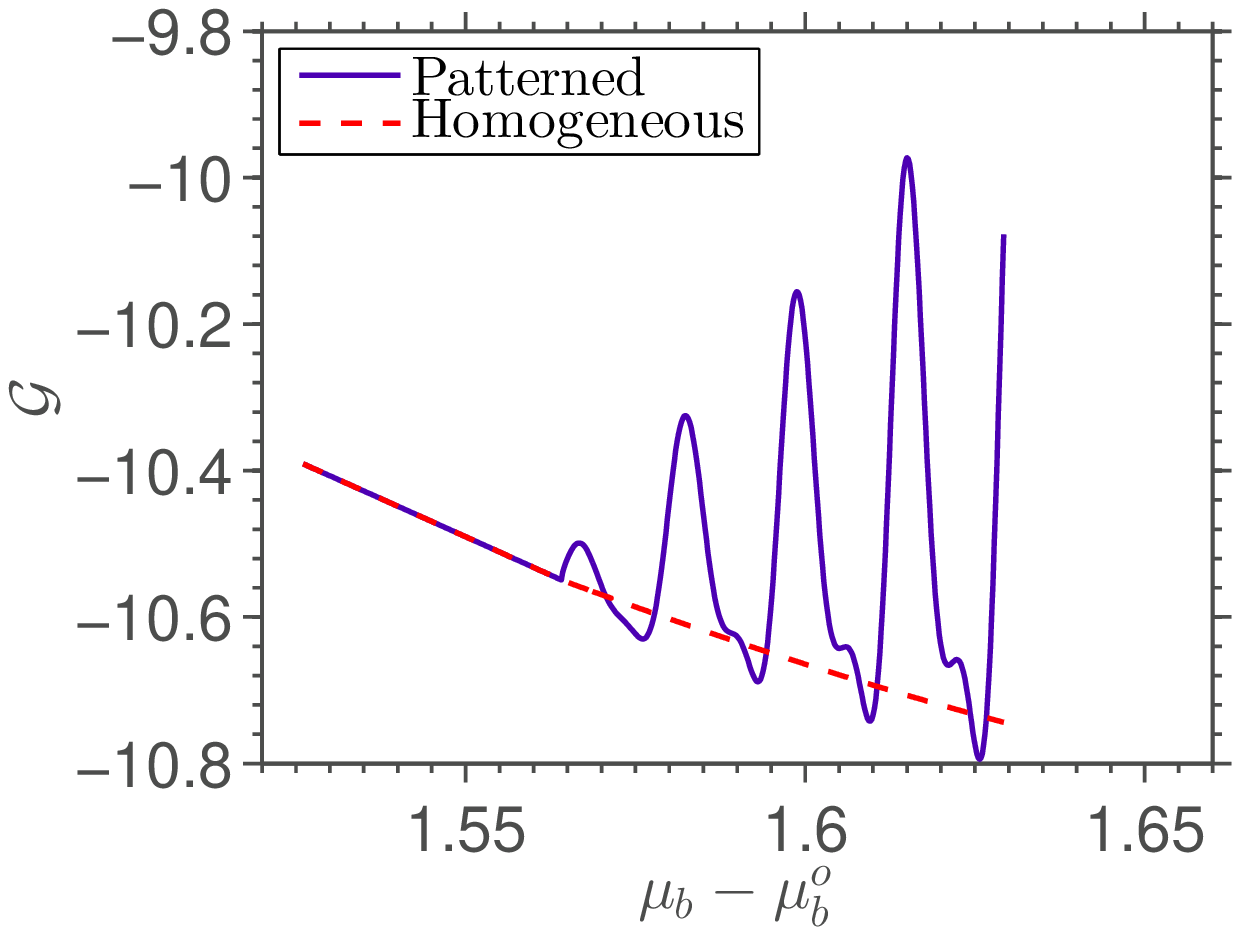}}\hfill
\subfigure[{\label{fig:slope}}]{\includegraphics[width=0.415\textwidth]{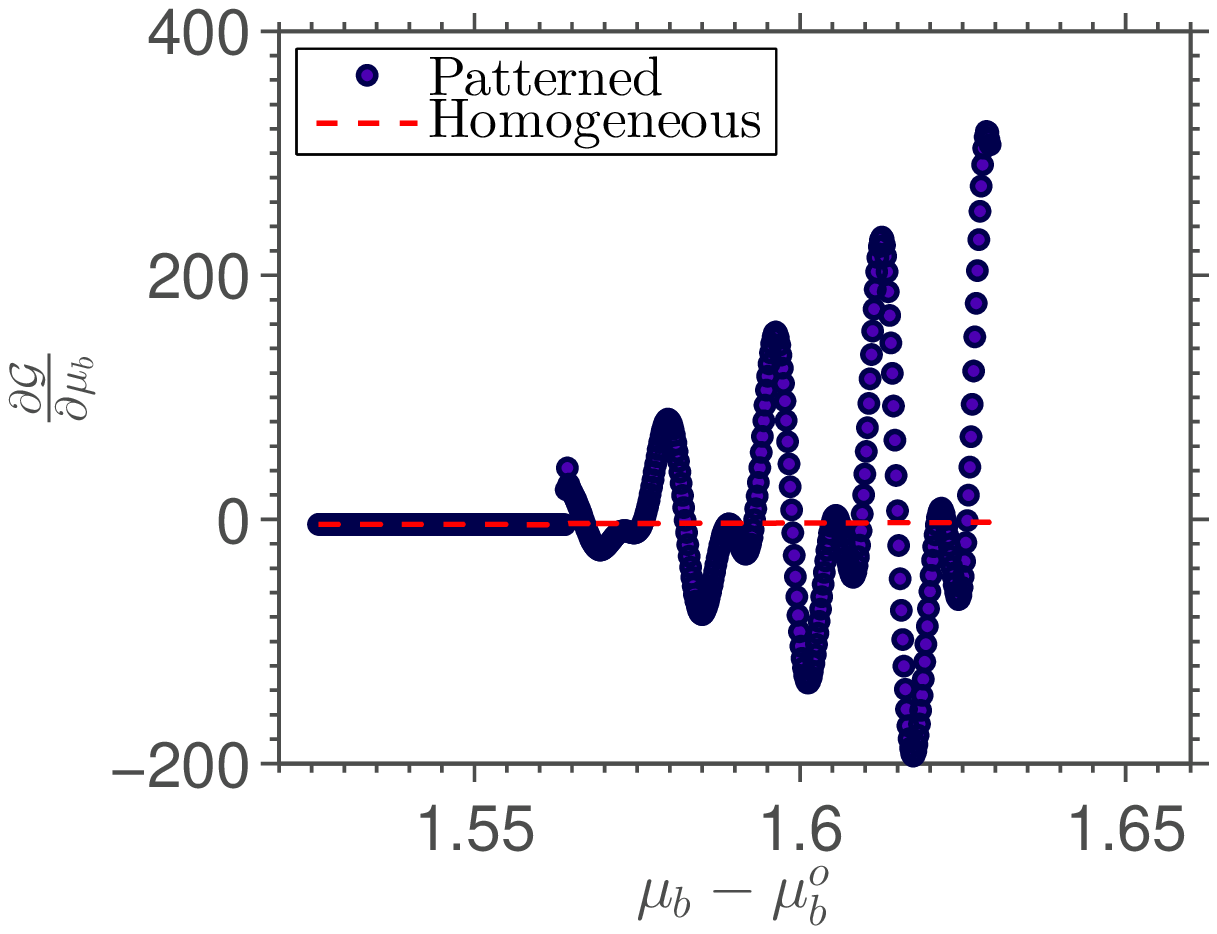}}
\subfigure[{\label{fig:sggh}}]{\includegraphics[width=0.415\textwidth]{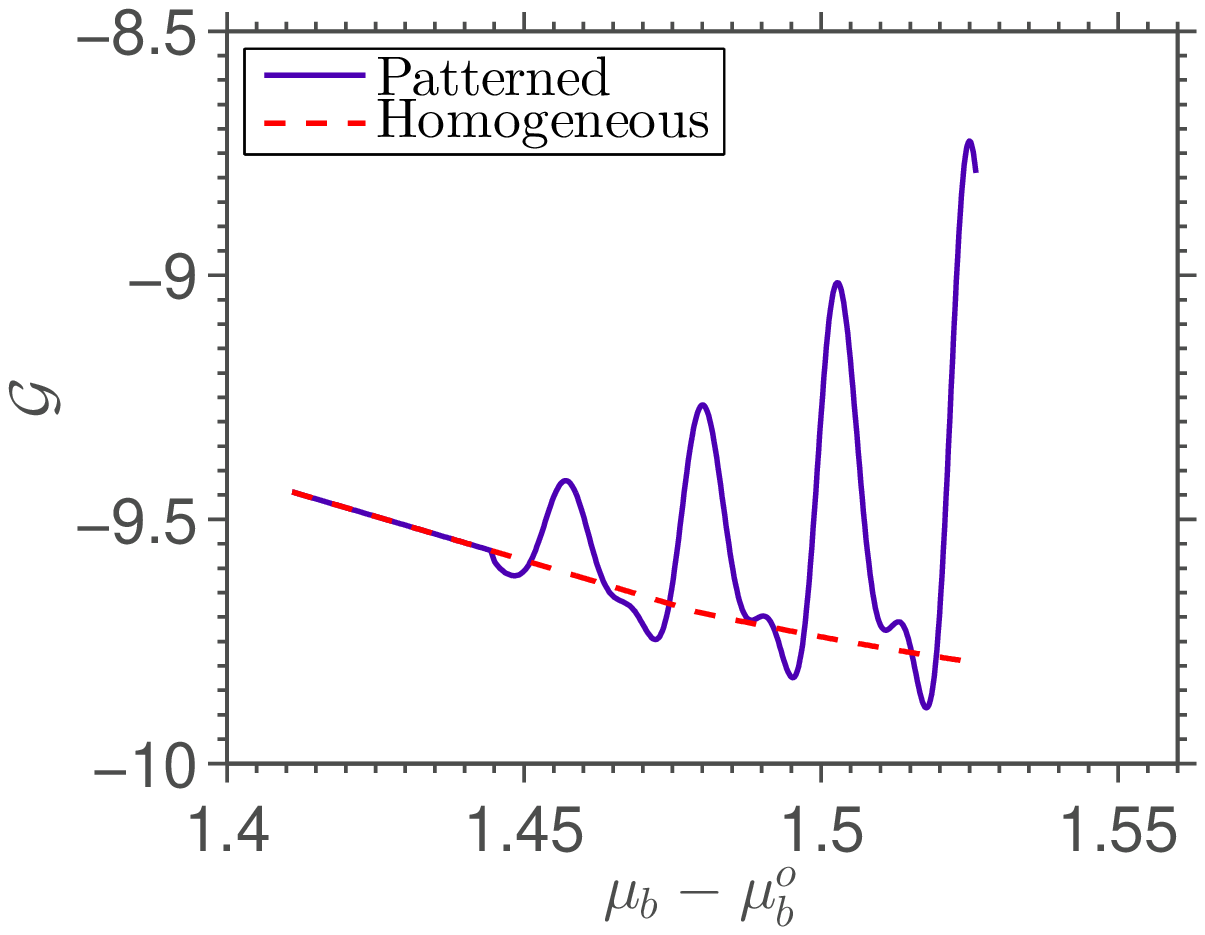}}\hfill
\subfigure[{\label{fig:slopeh}}]{\includegraphics[width=0.415\textwidth]{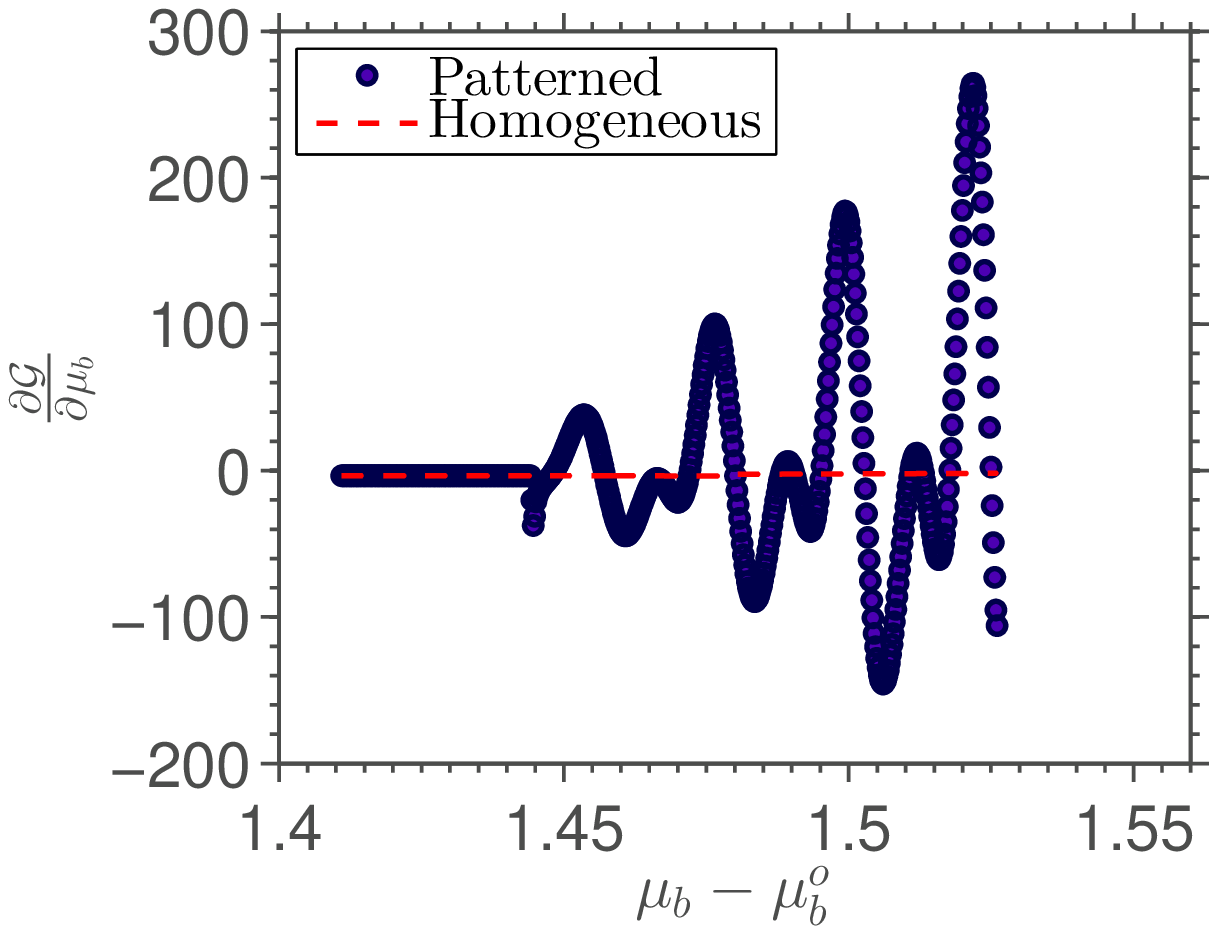}}
\caption{\label{fig:sggslope}Semi Grand Gibbs(SGG) free energy(left column) and corresponding slope(right column) profiles calculated analytically as function of chemical potential of control parameter, $b$ in 1D the Brusselator model at $t=150$ and $T=300K$ for three different values of parameter, $a$ leading to three different scenarios of Turing-Hopf interplay: FIG. \ref{fig:sggt} and \ref{fig:slopet} for $a=2.1$ Turing first; FIG. \ref{fig:sgg} and \ref{fig:slope} for COD2 Turing and Hopf appear simultaneously; FIG. \ref{fig:sggh} and \ref{fig:slopeh} for $a=1.8$  Hopf fist.  $A$ and $B$ are considered as the reference chemostatted species to define nonequilibrium free energy in open system. The dotted lines are for unstable homogeneous state of the system with no pattern. For all the cases diffusion  coefficients are: $D_{11}=D_{22}=1; D_{12}=0.51; D_{21}=-0.51$ and reaction rate constants are $K_{-\rho}=10 ^{-4} << K_{\rho}=1.$}
\end{figure*}

Analytical concentration field of intemediate species, $X$ as a function of control parameter, $b$ due to Turing-Hopf interplay as shown in first two row of FIG. \ref{compar} is calculated by using eq. \eqref{wave} in sec. \ref{thinteraction}. Corresponding reaction and diffusion entropy production rate is obtained from eq. \eqref{eprr} and \eqref{eprdd}, respectively in presence of cross diffusion. Study of entropy production rate separately for diffusion and reaction reveals  their proper contribution to total entropy production. It also renders clearly how Turing-Hopf interplay and cross diffusion, modify these two parts separately. In FIG. \ref{fig:EPBt}, as parameter value reaches  Turing instability critical point, nonzero entropy production rate due to both reaction and diffusion shows initially dynamical bifurcation kind of characteristics. Then appearance of Hopf instability modifies reaction part of the entropy production rate by its innate limit cycle type oscillatory dynamics and give rise to irregular oscillatory response of entropy production rate  with respect to parameter, $b$. Dotted red line in the same figure also shows modification of the diffusive entropy production rate by Hopf instability to a very little but finite extent and thus hints that limit cycle of Hopf instability has indirect dependence on the diffusion. In FIG. \ref{fig:EPBh}, initially Hopf instability is only present in  reaction diffusion system and entropy production rate is zero. Then as control parameter changes and exceeds the Turing critical point, a nonzero diffusion entropy production rate appears due to Turing instability. This means thermodynamic entity is modified by the Turing instability in this framework although Hopf instability appears first. It is an interesting result as in dynamical framework, Hopf instability screens the Turing instability if the former precedes the later instability in  reaction diffusion system. In FIG. \ref{fig:EPB} we can see as the Turing and Hopf instability appears simultaneously, both reaction and diffusion entropy production rate is modified sufficiently  from the initial zero value. Corresponding concentration profiles in all the cases are shown in first row of  FIG. \ref{compar}.

The left column of FIG. \ref{fig:sggslope} shows the semigrand Gibbs free energy change as a function of chemical energy of the control parameter, $b$. As suggested by the FIG. \ref{fig:sggt},\ref{fig:sgg} and \ref{fig:sggh}for 'Turing first', 'Co-dimension 2(COD2)' and 'Hopf first' respectively, the  transformed Gibbs free energy of unstable homogeneous part basically set the baseline for the transformed Gibbs free energy corresponding to the part where pattern formation arises. This clearly suggests the transformed Gibbs free energy plays the role of proper nonequilibrium thermodynamic potential of  reaction diffusion system in the presence of Turing-Hopf interplay, at least in global sense. Plot of slopes for the same thermodynamic entity is shown in the right column of FIG. \ref{fig:sggslope} to get more clear idea about the phase transitions in the response of the thermodynamic entity for whether Turing(FIG. \ref{fig:slopet}) or Hopf instability(\ref{fig:slopeh}) appears first or both of them appear simultaneously(FIG. \ref{fig:slope}) as control parameter, $b$ is varied. In FIG. \ref{compar} and \ref{fig:sggslope}, due to Turing Hopf interplay one obtains oscillation in concentration and thermodynamic quantities with $b$ and $\mu_b$. A series of phase transitions can open up the opportunity to control concentration and free energy profile both spatially and dynamically with varying the chemostatted species, $B$.

\section{\label{con}Conclusions}

In this work, we have investigated energetic and entropic cost of pattern arising in the realm of Turing-Hopf interplay in a standard model system by determining proper nonequilibrium potential and entropy production rate in  open system with finite size. In a systematic way we have shown here how the concentration, nonequilibrium semi grand Gibbs free energy and entropy production rate at steady state drastically depend on a control parameter in the Turing-Hopf interplay regime for three possible situations. This approach will also help to control and manipulate the efficiency and dissipation of a system far away from equilibrium. It also paves the way to relate Turing-Hopf interplay with the instance of nonequilibrium phase transitions which  generates  a possibility of huge modulation of free energy and concentration profiles. Here, we capture as well as quantify the effect of diffusion on the Hopf limit cycle through the diffusion-driven Turing instability. Proportionality of total EPR with the global concentration profile is an important result in the context of entropic cost of pattern formation and thus for the evolution of real chemical or biological systems in more larger sense. Furthermore, we have found that these outcomes are also valid for the experimentally found magnitudes of the self and cross diffusion coefficients. The only thing that would be different for this experimental values of  diffusion coefficients is the period of oscillations as the parameter $'a'$ on which critical frequency of the Hopf bifurcation depends, is shifted to new value due to this different set of the diffusion coefficients.

Amplitude equation formalism, a universal description in terms of dynamical symmetry breaking near a bifurcation point, has been utilized here to lay the basis of analytical construction. 
Approximate amplitude solution obtained by the analytical scheme for equal self-diffusion coefficients can describe the dynamical phenomena found in the experiments with high accuracy\citep{Lavrova2009PhaseInflux}. In our approach of finding the amplitude equation by exploiting KB scheme, we have considered both the self and cross-diffusion coefficients which are generally not equal. In this aspect, our results related to amplitude equation is more general and would be quite useful in the environment where cross-diffusion is present.

It turns out that even in the absence of so-called ‘local activation and long-range inhibition’ condition\citep{Murray2003MathematicalApplications} with equal self-diffusion coefficients of the species, proper choice of cross-diffusion coefficients can lead to the diffusive instabilities as the mathematical expressions of the intrinsic critical values of control parameter as well as wave number explicitly contain cross-diffusion coefficients. So Turing instabilities considered here is essentially cross-diffusion driven and this kind of thermodynamic description is valid beyond the traditional Turing pattern. Our selection of the Brusselator model in this study excludes the possibility of subcritical Hopf bifurcation. Here we have inspected weak Turing-Hopf interplay and have not considered subharmonic  oscillation in this kind of interplay. We believe this framework for Turing-Hopf interplay will also be applicable to study the thermodynamics of Turing-Hopf interaction in superdiffusive two species model\citep{Tzou2009InteractionModel}.

This analytically tractable thermodynamic description of the reaction-diffusion system is found to be powerful enough to capture almost all the essential richness of Turing-Hopf interaction in open chemical network. 
In linear Nonequilibrium thermodynamics, the thermodynamic driving force is specified as flux times Onsager coefficients near equilibrium. Here reaction affinity for a system kept far away from equilibrium is expressed directly from the elementary chemical reaction containing the nonlinear autocatalytic reaction. This approach of nonequilibrium thermodynamics on top of nonlinear dynamical features considered here for pattern formation could also be implemented in kinetic proofreading\citep{Hopfield1974KineticSpecificity,  2012StochasticBiophysics}, enzyme assisted copolymerization\citep{Andrieux2008NonequilibriumProcesses.} and in several nonequilibrium steady states of biochemical systems\citep{qian2006open}.
\bibliography{turing-hopf-ther}
\end{document}